\documentclass[reprint,superscriptaddress,amsmath,amssymb,a4paperaps,pra,nofootinbib]{revtex4-1}
\usepackage{dcolumn,amssymb,amsmath,amsfonts,graphicx,latexsym,float,xcolor}
\usepackage{epstopdf}
\usepackage{nameref}
\usepackage{txfonts}
\usepackage{ulem}
\usepackage{easyReview}
\begin{document}

\title{Asymptotic population imbalance of an ultracold bosonic ensemble in a driven double-well }
\author{Jie Chen}
\email {jie.chen@physnet.uni-hamburg.de}
\affiliation{Zentrum f\"ur Optische Quantentechnologien, Fachbereich Physik, Universit\"at Hamburg, Luruper Chaussee 149, 22761 Hamburg, Germany}
\author{Aritra K. Mukhopadhyay}
\email {aritra.mukhopadhyay@physnet.uni-hamburg.de}
\affiliation{Zentrum f\"ur Optische Quantentechnologien, Fachbereich Physik, Universit\"at Hamburg, Luruper Chaussee 149, 22761 Hamburg, Germany}
\author{Peter Schmelcher}
\email{pschmelc@physnet.uni-hamburg.de}
\affiliation{Zentrum f\"ur Optische Quantentechnologien, Fachbereich Physik, Universit\"at Hamburg, Luruper Chaussee 149, 22761 Hamburg, Germany}
\affiliation{The Hamburg Centre for Ultrafast Imaging, Universit\"at Hamburg, Luruper Chaussee 149, 22761 Hamburg, Germany}
\date{\today}

\begin{abstract}
We demonstrate that an ultracold many-body bosonic ensemble confined in an one-dimensional (1D) double well potential exhibits a population imbalance between the two wells at large timescales, when the depth of the wells are modulated by a time-dependent driving force. The specific form of the driving force is shown to break spatial parity and time-reversal symmetries, which leads to such an asymptotic population imbalance (API). The value of the API can be flexibly controlled by changing the phase of the driving force and the total number of particles. While the API is highly sensitive to the initial state in the few-particle regime, this dependence on the initial state is lost as we approach the classical limit of large particle numbers. We perform a Floquet analysis in the few-particle regime and an analysis based on a driven classical non-rigid pendulum in the many-particle regime. Although the obtained API values in the many-particle regime agree very well with that obtained in the classical limit, we show that there exists a significant disagreement in the corresponding real-time population imbalance due to quantum correlations.
\end{abstract}

\maketitle
\section{Introduction}
Ultracold atomic gases provide an ideal platform for the study of quantum many-body physics \cite{cold_atom_rev}. Ever since the realization of Bose–Einstein condensates of weakly interacting gases  \cite{BEC_1,BEC_2}, milestone achievements have been reported in cold-atom experiments. Prominent examples are the observation of the superfluid to Mott insulator phase transition of bosons in optical lattices \cite{BH_exp_1,BH_exp_2,BH_exp_3} and the BCS-BEC crossover for a degenerate Fermi gases \cite{mixture_exp_ff_1, mixture_exp_ff_2}. Among them, trapping of bosonic atoms in a double-well potential constitutes a prototype system for the investigations of the tunneling dynamics  \cite{DW_exp_1,DW_exp_2, DW_exp_3}. Such a system represents a bosonic Josephson junction (BJJ), an atomic analogy of the Josephson effect initially predicted for a pair of electrons (Cooper pair) tunneling through two weakly linked superconductors  \cite{BJJ_1,BJJ_2}. Owing to the unprecedented controllability of the trapping geometries as well as the atomic interaction strengths \cite{cold_atom_rev}, studies of the BJJ unveil various intriguing phenomena which are not accessible for conventional superconducting systems \cite{BJJ_Rabi_1, BJJ_Rabi_2, BJJ_Rabi_3, BJJ_Frag_1,BJJ_Frag_2, BJJ_Squeeze_1, BJJ_Squeeze_2, BJJ_Few_1, BJJ_Few_2, BJJ_Few_3, BJJ_Few_4, BJJ_Few_5}. Examples are the Josephson oscillations \cite{BJJ_Rabi_1, BJJ_Rabi_2, BJJ_Rabi_3}, fragmentations  \cite{BJJ_Frag_1, BJJ_Frag_2}, macroscopic quantum self trapping \cite{DW_exp_3, BJJ_Rabi_1, BJJ_Rabi_2},  collapse and revival sequences \cite{BJJ_Rabi_3}, atomic squeezing state \cite{BJJ_Squeeze_1, BJJ_Squeeze_2} as well as strongly correlated tunneling dynamics in few-body systems \cite{BJJ_Few_1, BJJ_Few_2, BJJ_Few_3, BJJ_Few_4, BJJ_Few_5}. 

On the other hand, systems driven out of equilibrium by time dependent driving forces have attracted growing interests in recent years. The time-dependent variation of the control parameters can trigger non-trivial responses allowing the system to exhibit novel properties which are absent in the static counterpart \cite{Driven_rev_1, Driven_rev_2, Driven_rev_3}. It has been shown that external driving can lead to different phenomena in ultracold atomic ensembles \cite{Driven_rev_3}, for instance, the emergence of superfluid-Mott insulator transition by periodically shaking the optical lattices \cite{Driven_BEC_Mott_1,Driven_BEC_Mott_2,Driven_BEC_Mott_3}, the single-particle and many-body coherent destruction of tunneling in a driven double-well potential \cite{CDT_1,CDT_2}. A phenomena of particular interest in driven cold atomic ensembles is the `ratchet effect', which can lead to an unidirectional transport of the atoms in a fluctuating environment even in absence of a net force bias \cite{Ratchet_1, Ratchet_2, Ratchet_3, Ratchet_4}. In order to realize such directed transport, the system must necessarily break certain spatio-temporal symmetries \cite{Ratchet_rev, Ratchet_5, Ratchet_6, Ratchet_7, Ratchet_8, Ratchet_9, Ratchet_10, Ratchet_11, Ratchet_12}. This provides not only a useful method for controlling the transport of atomic ensembles but also different applications like particle separation based on physical properties \cite{Ratchet_13, Ratchet_14, Ratchet_15} and design of efficient velocity filters \cite{Ratchet_16, Ratchet_17}.

In the present work, we explore the ratchet effect for a many-body bosonic ensemble confined in a 1D double-well potential whose depth is periodically modulated. Unlike most previous studies, which focus either on the non-interacting regime  \cite{CDT_1} or on the transient dynamics \cite{BJJ_Driven_1, BJJ_Driven_2}, we investigate the transport properties of interacting particles in the asymptotic limit $t \rightarrow \infty$. Specifically, we start with an equal number of particles in both wells and explore the emergence of an asymptotic population imbalance (API) of particles in the two wells. For this, the spatial parity and the time reversal symmetries need to be broken \cite{Ratchet_rev, Ratchet_5}, which is achieved by a suitable bi-harmonic driving force. We show that the value of the API can be flexibly controlled by changing the driving phase. Most importantly, we demonstrate that for the same driving force, the value of the API shows an individually characteristic behavior for different particle numbers. While the API is highly sensitive to the initial state in the few-particle regime, this dependence on the initial state is lost as the number of particles is increased thus approaching the classical limit of large particle numbers. We explain the behavior of the API in the few particle limit in terms of the underlying Floquet modes. In the many-particle regime, we show that the API can be interpreted in terms of the well established classical non-rigid driven pendulum \cite{BJJ_Rabi_1, BJJ_Rabi_2, BJJ_Rabi_3, BJJ_Driven_1}, providing a deeper insight into the connections between classical and quantum physics. Although the obtained API values agree very well with the ones in the classical limit, we show that there exists a significant disagreement in the corresponding real-time population imbalance due to the presence of quantum correlations.

This paper is organized as follows. In Sec. \ref{Hamiltonian_setup}, we introduce our setup and the quantities of interests. In Sec. \ref{symmetries}, we investigate the relevant symmetries controlling the API in both the quantum and classical limits. In Sec. \ref{Results_analysis} and Sec.\ref{discussion}, we present a comprehensive study of the behavior of the API as we go from the few-particle regime to the many-particle regime. Finally, our conclusions and outlook are provided in Sec. \ref{Conclusions}.  

\section{Setup} \label{Hamiltonian_setup}
We consider an ultracold many-body ensemble consisting of $N$ interacting bosons confined within an one dimensional (1D) symmetric double-well potential $V_{DW}(x)$, whose depth is modulated periodically via a driving force $F(t)$. The Hamiltonian of the system is given by 
\begin{align}
\hat{H}(t) &=\int dx~\hat{\psi}^{\dagger}(x) \textit{h}_{0}(x,t) \hat{\psi}(x) \nonumber\\
&+\frac{g_{b}}{2}\int dx~\hat{\psi}^{\dagger}(x)\hat{\psi}^{\dagger}(x)\hat{\psi}(x)\hat{\psi}(x),
\end{align}
where $ \hat{\psi}^{\dagger}(x)$ [$\hat{\psi}(x)$] is the field operator that creates (annihilates) a boson at position $x$. The single-particle Hamiltonian $\textit{h}_{0}(x,t) = -\frac{\hbar^{2}}{2 m}\frac{\partial^{2}}{\partial x^{2}}+ V_{DW} (x) + xf(t)$ , where $f(t) = E_{1}\text{cos}(\omega t) + E_{2}\text{cos}(2\omega t + \phi)$ is a bi-harmonic periodic driving force. $E_{1}$ and $E_{2}$ denote driving amplitudes, $\omega$ is the driving frequency and $\phi$ is a temporal phase shift. The interaction among the bosons is assumed to be of zero-range and is modeled by a contact potential of strength \cite{Feshbach_0,Feshbach_1}
\begin{equation}
g_{b} = \frac{4 \hbar^{2}a_{b}}{m a_{\bot, b}^{2}}[1-C \frac{a_{b}}{a_{\bot, b}}]^{-1}.
\end{equation}
Here $a_{b}$ is the 3D Bose-Bose $s$-wave scattering length and $C \approx 1.4603$ is a constant. The parameters $a_{\bot, b} = \sqrt{2 \hbar/ \omega_{\bot}}$ describes the transverse confinement. In this work, we focus on the repulsive interaction regime, i.e., $g_{b} \geqslant 0$, which can be controlled experimentally by tuning the $s$-wave scattering lengths via Feshbach or confinement-induced resonances \cite{Feshbach_1,Feshbach_2,Feshbach_3}.

For sufficiently weak interaction and tight enough confinement, the particle excitations are severely suppressed and as a result, the bosons mainly populate the lowest two eigenstates $u_{\pm} (x) $ for the single-particle Hamiltonian $\hat{h}_{s}(x) = -\frac{\hbar^{2}}{2 m}\frac{\partial^{2}}{\partial x^{2}}+ V_{DW} (x) $. We, therefore, adopt the single-band approximation by expanding the field operator as 
\begin{equation}
\hat{\psi}(x) = u_{L}(x) \hat{a}_{L} + u_{R}(x) \hat{a}_{R}, \label{2_mode_psi}
\end{equation}
with $u_{L,R}(x)$ being the Wannier-like states localized in the left and right well, respectively. This leads to the modified Hamiltonian 
\begin{align}
\hat{H}_{BH}(t) &=  -J_{BH} (\hat{a}^{\dagger}_{L}\hat{a}_{R} + \hat{a}^{\dagger}_{R} \hat{a}_{L} ) + \frac{U_{BH}}{2} \sum_{i = L,R} \hat{a}^{\dagger}_{i}\hat{a}^{\dagger}_{i}\hat{a}_{i} \hat{a}_{i} \nonumber \\
&+ f(t) (\hat{a}^{\dagger}_{L} \hat{a}_{L}  - \hat{a}^{\dagger}_{R} \hat{a}_{R}). \label{BH_model}
\end{align}
corresponding to the two-site Bose-Hubbard (BH) model with $\hat{a}^{\dagger}_{L/R}$ ($ \hat{a}_{L/R}$) being the creation (annihilation) operator with respect to the $u_{L/R}(x)$ state. The coefficients
\begin{align}
J_{BH} &= \int u_{L}^{\ast}(x) h_{s}(x)  u_{R}^{\ast}(x), \nonumber \\
U_{BH} &= g_{b} \int u_{i}^{4}(x)  dx, ~~~ (i = L,R) 
\end{align}
represent the hopping amplitude and the on-site repulsion energy, respectively, and 
\begin{equation}
f(t) =  E_{1}\text{cos}(\omega t) + E_{2}\text{cos}(2\omega t + \phi) \label{f_t}
\end{equation}
denotes the bi-harmonic driving force. We choose the units of the energy and time as $\eta = \epsilon_2 - \epsilon_1$ and $\xi = 2\pi \hbar / \eta$, with $\epsilon_1$ ($\epsilon_2$) being the energy of the ground (first excited) state of the single-particle Hamiltonian $\hat{h}_{s}(x)$. With this choice, the hopping amplitude in $\hat{H}_{BH}(t)$ results in a constant value $J_{BH} = 1/2$.

In this work, we explore the asymptotic particle transport in the setup due to the time-dependent driving of the spatial potential. Since our system is spatially bounded, such a particle transport eventually results in an \textit{asymptotic population imbalance} (API) between the two wells. We characterize the API as 
\begin{equation}
\overline{\Delta \rho } = lim_{ \tau,  \tau^{\prime} \rightarrow \infty} ~\frac{1}{ \tau^{\prime}} \int_{\tau}^{\tau + \tau^{\prime}} dt~ \langle \Delta \hat{\rho}\rangle(t),  \label{API}
\end{equation}
with $ \Delta \hat{\rho} = (\hat{n}_L - \hat{n}_R)/N$ being the normalized particle occupation difference for a fixed total particle number $N$. The average $\langle \Delta \hat{\rho}\rangle(t)$ is computed with respect to the many-body wavefunction $| \Psi(t) \rangle$, which evolves according to the Schr\"odinger equation $i\hbar \partial / \partial t | \Psi(t) \rangle = \hat{H}_{BH}(t) |\Psi(t) \rangle $. Throughout this work, we consider the initial population of the two wells to be equal such that $\langle \Delta \hat{\rho}\rangle(0)= 0 $ (see below), and we explore the possibilities for the appearance of a non-vanishing $\overline{\Delta \rho } $ in the limit $ \tau, \tau^{\prime} \rightarrow \infty$.

Since the Hamiltonian \eqref{BH_model} is periodic in time, i.e., $H_{BH}(t) = H_{BH}(t+T)$, with period $T = 2\pi/ \omega$, we can write the above wavefunction as \cite{Driven_rev_1,Floquet_1}
\begin{equation}
|\Psi(t) \rangle = \sum_{\alpha} A_{\alpha} e^{-i \epsilon_{\alpha}t} | \Phi_{\alpha}(t) \rangle,
\end{equation}
with $| \Phi_{\alpha}(t) \rangle$ being the Floquet mode (FM) with the temporal period $T$, i.e., $| \Phi_{\alpha}(t) \rangle = | \Phi_{\alpha}(t+ T) \rangle$. The quasi-energy (QE) $\epsilon_{\alpha}$ can always be chosen within the interval $[-\omega/2, \omega/2]$ \cite{Driven_rev_1,Floquet_1}. According to the Floquet theorem, the FM fulfills the eigenstate equation
\begin{equation}
\hat{H}_{F}(t) | \Phi_{\alpha}(t) \rangle \rangle = \epsilon_{\alpha} |\Phi_{\alpha}(t) \rangle \rangle. \label{Floquet_eigen}
\end{equation}
Here $\hat{H}_{F}(t) = \hat{H}_{BH}(t) - i \partial / \partial t$ is the Floquet Hamiltonian which is defined in the composite Hilbert space $\mathcal{R} \otimes \mathcal{T}$, with $\mathcal{R}$ being the Hilbert space of square integrable functions and $\mathcal{T}$ denotes the space of time-periodic functions whose period is $T = 2\pi/ \omega$. The FM $|\Phi_{\alpha}(t) \rangle \rangle$ can thus be expressed as a linear superposition of the composite states 
\begin{equation}
|\Phi_{\alpha}(t) \rangle \rangle = \sum_{N_{L}, n} \mathcal{D}^{\alpha}_{N_{L}, n} |N_{L}, N_{R}\rangle \otimes e^{i n \omega t}, \label{FM_number_state}
\end{equation}
where $\{|N_{L}, N_{R}\rangle \}$ denote the number states with $N_{L} + N_{R} = N$ and $n =0, \pm 1, \pm 2, ...$ is an integer number. Correspondingly, the orthonormality condition for FMs read
\begin{align}
\langle \langle \Phi_{\alpha}(t) |  \Phi_{\beta}(t) \rangle \rangle &= \sum_{N_{L}, n} \sum_{N_{L}^{\prime}, n^{\prime}} [\mathcal{D}^{\alpha}_{N_{L}, n}]^{*} \mathcal{D}^{\beta}_{N_{L}^{\prime}, n^{\prime}} \nonumber \\
&\times \frac{1}{T} \int_{0}^{T} dt e^{i(n^{\prime}-n) \omega t} \langle N_{L}, N_{R} |N_{L}^{\prime}, N_{R}^{\prime} \rangle = \delta_{\alpha, \beta}
\end{align}

In terms of the Floquet modes, the API defined in Eq.\eqref{API} simplifies to 
\begin{align}
\overline{\Delta \rho} & = lim_{\tau,  \tau^{\prime} \rightarrow \infty} ~\frac{1}{ \tau^{\prime} } \int_{\tau}^{\tau +  \tau^{\prime}} dt~ \sum_{\alpha, \beta} A_{\alpha}^{*} A_{\beta} ~e^{i(\epsilon_{\alpha} - \epsilon_{\beta})t} \langle \langle\Phi_{\alpha}(t)| \Delta \hat{\rho} | \Phi_{\beta}(t)\rangle \rangle \nonumber \\
& = \sum_{\alpha} P_{\alpha} \overline{\Delta \rho_{\alpha}},  \label{API_FM}
\end{align}
where $P_{\alpha} = A_{\alpha}^{*} A_{\alpha} $ denotes the weight corresponding to the $\alpha$-th FM and is obtained as the overlap of the initial state with the $| \Phi_{\alpha}(t=0) \rangle \rangle$. $ \overline{\Delta \rho_{\alpha}} =  \langle \langle \Phi_{\alpha}(t)| \Delta \hat{\rho} | \Phi_{\alpha}(t)\rangle \rangle $ denotes the API corresponding to the $\alpha$-th FM $| \Phi_{\alpha}(t) \rangle \rangle$. It is important to emphasize that the validity for the Eq. \eqref{API_FM} relies on the assumption that the Floquet Hamiltonian $\hat{H}_{F}(t)$ is non-degenerate, i.e., $\epsilon_{\alpha} \neq \epsilon_{\beta}$ for $\alpha \neq \beta$, which is well-justified by the extension of the von Neumann-Wigner theorem \cite{ Ratchet_rev, Non_degenerate}.

\section{Symmetry analysis}\label{symmetries}
In order to achieve a non-vanishing asymptotic population imbalance between the two wells, one needs to break certain symmetries of the underlying system, specifically the \textit{generalized parity symmetry} and the \textit{generalized time-reversal symmetry}. In this section, we discuss how these symmetries are violated in our system for both the quantum and the classical cases. We begin with the quantum limit where we show how these symmetries affect both the FMs $|\Phi_{\alpha}(t) \rangle \rangle$ and the operator $ \Delta \hat{\rho}$, thereby controlling the value of the API. In contrast, the dynamics of the particles in the classical limit is fully characterized by the classical phase space. The appearance of a nonzero API in this case, as we will show, is due to a desymmetrization of the chaotic manifold of the phase space caused by the breaking of the symmetries.

\subsection{Quantum limit}
\subsubsection{Angular-momentum representation}
We first introduce three angular-momentum operators as \cite{BJJ_Rabi_3, BJJ_Driven_1}
\begin{align}
\hat{J}_{x} &= \frac{1}{2} (\hat{a}_{L}^{\dagger} \hat{a}_{R}  + \hat{a}_{R}^{\dagger} \hat{a}_{L} ) ,~~~\hat{J}_{y} = -\frac{i}{2} (\hat{a}_{L}^{\dagger} \hat{a}_{R}  - \hat{a}_{R}^{\dagger} \hat{a}_{L} ), \nonumber\\
\hat{J}_{z} &= \frac{1}{2} (\hat{a}_{L}^{\dagger} \hat{a}_{L}  - \hat{a}_{R}^{\dagger} \hat{a}_{R} ).  \label{spin_operators}
\end{align}
obeying the SU(2) commutation relation $ [\hat{J}_{\alpha}, \hat{J}_{\beta}] = i \epsilon_{\alpha \beta \gamma} \hat{J}_{\gamma}$. In this representation, the many-particle Hamiltonian \eqref{BH_model} can be rewritten as 
\begin{equation}
\hat{H}_{S}(t) = - \hat{J}_{x} + U_{BH} \hat{J}_{z}^{2} -2f(t) \hat{J}_{z}, \label{spin_BH_model}
\end{equation}
and the Floquet Hamiltonian in Eq.\eqref{Floquet_eigen} becomes as $\hat{H}_{F}(t) = \hat{H}_{S}(t) - i \partial / \partial t$. The Casimir invariant $\hat{J}^{2}$ can be expressed in terms of the total number of particles $N$ as
\begin{equation}
\hat{J}^{2} = \hat{J}_{x}^{2} + \hat{J}_{y}^{2}  + \hat{J}_{z}^{2} = \frac{N}{2}(\frac{N}{2}+1),
\end{equation}
denoting the conservation of the total angular momentum with the magnitude $l = N/2$. Consequently, all the eigenstates $\{|l,m \rangle \}$ for both $\hat{J}^{2}$ and $\hat{J}_{z}$ precisely corresponds to the $N + 1$ basis states $\{|N_{L}, N_{R}\rangle \}$ of the $N$-particle Hilbert space. In this way, the original many-particle Hamiltonian in \eqref{BH_model} is completely mapped onto the single-particle Hamiltonian in \eqref{spin_BH_model}. The hopping of the particles between the two wells now corresponds to an angular momentum precession around about the $x$-axis and the driving potential $f(t) (\hat{N}_{L}  -  \hat{N}_{R})$ can be interpreted as a periodic modulation of a Zeeman field applied in the $z$-direction. The FMs in the Eq \eqref{FM_number_state} can be now expressed as
\begin{equation}
| \Phi_{\alpha}(t) \rangle \rangle = \sum_{m,n} C_{m,n}^{\alpha} | l,m \rangle  \otimes e^{i n \omega t}, \label{cof_floquet}
\end{equation}
in terms of the angular momentum basis $\{| l,m \rangle \} $. The API can hence be interpreted as the asymptotic magnetization along $z$-direction
\begin{align}
\overline{\Delta \rho } &=  \overline{J_{z} } =  lim_{\tau,  \tau^{\prime} \rightarrow \infty} ~\frac{2}{N \tau^{\prime} } \int_{\tau}^{\tau +  \tau^{\prime}}dt~ \langle \hat{J}_{z}\rangle(t) \nonumber \\
 &= \frac{2}{N} \sum_{\alpha} P_{\alpha} \langle \langle \Phi_{\alpha}(t)| \hat{J}_{z}  | \Phi_{\alpha}(t)\rangle \rangle =  \sum_{\alpha} P_{\alpha} \overline{J_{z}^{\alpha} }, \label{API_sum}
\end{align}
with $\overline{J_{z}^{\alpha} } =  \frac{2}{N}\langle \langle \Phi_{\alpha}(t)| \hat{J}_{z}  | \Phi_{\alpha}(t)\rangle \rangle$ being the API corresponding to the $\alpha$-th FM $| \Phi_{\alpha}(t)\rangle \rangle$. In order to obtain a non-zero API, it is important that the system breaks the symmetries which transforms $\hat{J}_{z} \rightarrow -\hat{J}_{z}$ and hence renders $\overline{J_{z}^{\alpha} }=0$ \cite{Ratchet_rev}. In the following we discuss the general form of these symmetry operations and how they can be broken.

\subsubsection{Generalized parity symmetry} \label{Sp_symmetry}

In the absence of any driving force, i.e. $ E_1= E_2 = 0$, the Hamiltonian in \eqref{spin_BH_model} is time-independent. A natural choice of the symmetry transformation which keeps this time-independent Hamiltonian invariant, meanwhile, changing the sign of $\hat{J}_{z}$ is a rotation through an angle $\pi$ about the $x$-axis denoted by the operator $\hat{R}_{x} (\pi)= e^{-i \pi \hat{J}_{x}}$. This is no longer true for the time dependent cases since $\hat{R}_{x}(\pi) \hat{H}_{F}(t) \hat{R}_{x} ^{-1}(\pi) \neq  \hat{H}_{F}(t) $ in general. However, if $ E_2 = 0$, the driving force changes sign due to a time translation, i.e. $f(t) = -f(t + T/2)$ [c.f. Eq.\eqref{f_t}], the Hamiltonian $\hat{H}_{F}(t)$ is symmetric with respect to the transformation \cite{kicked_top}
\begin{equation}
S_{p}:  (J_{x}, J_{z}, t)  \rightarrow (J_{x}, -J_{z}, t+T/2 ),
\end{equation}
generated by the symmetry operator
\begin{equation}
\hat{S}_{p} = \hat{R}_{x}(\pi) \otimes \hat{Q}(T/2).
\end{equation}
Here $\hat{Q}(T/2)$ is the time-shift operator which shifts $t$ by $T/2$, resulting in $f(t) \rightarrow -f(t)$. $\hat{S}_{p}$ is the most general transformation which keeps $\hat{H}_{F}(t)$ invariant but changes the sign of $\hat{J}_{z}$ in the presence of our periodic driving force $f(t)$. In view of the interpretation of $\hat{J}_{z}$ [c.f. Eq.\eqref{spin_operators}] in terms of the particle numbers in the left and right well of our double well potential, we regard the symmetry transformation $S_{p}$ as the \textit{generalized parity symmetry}. 

Since $\hat{S}_{p} \hat{H}_{F}(t) \hat{S}_{p} ^{-1} =  \hat{H}_{F}(t) $ and $\hat{S}_{p}$ is an unitary operator, all the eigenstates of $\hat{H}_{F}(t)$ can be characterized as either symmetric or anti-symmetric with respect to $\hat{S}_{p}$, i.e., $\hat{S}_{p} | \Phi_{\alpha}(t) \rangle = \pm \sigma | \Phi_{\alpha}(t) \rangle$ with $\sigma = 1$  for $l = N/2 $ being the integers and $\sigma = i$ for $l $ being the half-integers. Along with the relation that $\hat{S}_{p} \hat{J}_{z}  \hat{S}_{p}^{-1} = -\hat{J}_{z}$, this implies
\begin{align}
\overline{J_{z}^{\alpha} } &=  \frac{2}{N} \langle \langle \Phi_{\alpha}(t)| \hat{J}_{z} | \Phi_{\alpha}(t)\rangle \rangle \nonumber \\
&= \frac{2}{N} \langle \langle \Phi_{\alpha}(t) |\hat{S}^{-1}_{p} \hat{S}_{p} \hat{J}_{z} \hat{S}^{-1}_{p} \hat{S}_{p} |\Phi_{\alpha}(t)\rangle \rangle \nonumber\\
&= - \frac{2}{N} \langle \langle \Phi_{\alpha}(t)| \hat{J}_{z} | \Phi_{\alpha}(t)\rangle \rangle =- \overline{J_{z}^{\alpha} }=0. \label{zero_API_parity}
\end{align}
Here we have employed the fact that $\hat{S}_{p}$ is an unitary operator which leads to $\langle \langle \Phi_{\alpha}(t) |\hat{S}^{-1}_{p} = \langle \langle \Phi_{\alpha}(t)| \hat{S}^{\dagger}_{p} = \pm \sigma^{\ast} \langle \langle \Phi_{\alpha}(t)| $. Since the contribution $\overline{J_{z}^{\alpha} }$ from each FM  $| \Phi_{\alpha}(t) \rangle$ to the API $\overline{J_{z} }$ vanishes, one concludes that $\overline{J_{z} } = \sum_{\alpha} P_{\alpha} \overline{J_{z}^{\alpha} } = 0$ for any arbitrary initial condition. As the above single harmonic driving force (i.e. $E_{2} = 0$) satisfy $f(t) = -f(t + T/2)$, the corresponding API is always zero. Hence in order to achieve a non-zero API, we must have $E_{2} \neq 0$.

\subsubsection{Generalized time-reversal symmetry} \label{St_symmetry}

Apart from ${S}_{p}$, also the time reversal operation can flip the sign of $\hat{J}_{z}$ \cite{QM_Sakurai}.  In fact, for our bi-harmonic driving force with the temporal phase shift $\phi=\pi/2$ or $\phi = 3\pi/2 $, $f(t)$ satisfies $f(t) = -f(-t + T/2)$ [c.f. Eq.\eqref{f_t}], one can define the \textit{generalized time-reversal symmetry} transformation \cite{kicked_top}
\begin{equation}
S_{t}:  (J_{x}, J_{z}, t)  \rightarrow (J_{x}, -J_{z}, -t+T/2)
\end{equation}
generated by the symmetry operator
\begin{equation}
\hat{S}_{t} = \hat{R}_{z}(\pi) \otimes \hat{\Theta} \otimes \hat{Q}(T/2) \label{S_t}
\end{equation}
as the most general form of the time reversal operation which transforms $\hat{J}_{z} \rightarrow - \hat{J}_{z}$ and keeps the Hamiltonian $\hat{H}_{F}(t)$ invariant. Here $\hat{R}_{z}(\pi)$ represents the operator inducing a rotation by an angle $\pi$ around the $z$-axis, $\hat{\Theta}$ is the anti-unitary time-reversal operator and $\hat{Q}(T/2)$ is the time-shift operator. Although the time-reversal operator $\hat{\Theta}$ does not commute with the time-shift operator $\hat{Q}(T/2)$ in general, we note that, when they are acting on the FMs, the relative order among them does not affect the physics (see Appendix \ref{Appendix_1}).

Due to the anti-unitary operator $\hat{\Theta}$ in $\hat{S}_{t}$, one cannot classify the FMs based on odd or even symmetry analogous to our previous discussion for the parity transformation. However, we note that \cite{QM_Sakurai}
\begin{align}
\hat{R}_{z}(\pi) | l,m \rangle  &= e^{-im\pi} |l,m \rangle , \nonumber \\ 
\hat{\Theta} | l,m \rangle & = i^{2m} |l,-m \rangle.
\end{align}
These relations together with the fact that the transformation $\hat{S}_{t}$ preserves the modulus of the inner product of two FMs provide an useful relation regarding the expansion coefficients
\begin{equation}
|C_{m,n}^{\alpha}|^{2} = |C_{-m,n}^{\alpha}|^{2}. \label{cof_TR_FM}
\end{equation}

We note that the API $\overline{J_{z}^{\alpha} }$ corresponding to the FM $| \Phi_{\alpha}(t)\rangle$ can be expressed in terms of the coefficients $C_{m,n}^{\alpha}$ as
\begin{equation}
\overline{J_{z}^{\alpha} } = \frac{2}{N} \langle \langle \Phi_{\alpha}(t)| \hat{J}_{z} | \Phi_{\alpha}(t)\rangle \rangle =  \frac{2}{N} \sum_{m,n} |C_{m,n}^{\alpha}|^{2} m. \label{cof_TR_FM2}
\end{equation}
Alternatively, by applying the symmetry transformation $\hat{S}_{t}$,  $\overline{J_{z}^{\alpha} }$ can also be expressed as \cite{QM_Sakurai}
\begin{align}
\overline{J_{z}^{\alpha} } &= \frac{2}{N}  \langle \langle \Phi_{\alpha}(t)|  \hat{J}_{z}  |\Phi_{\alpha}(t)\rangle \rangle \nonumber \\
& =  \frac{2}{N}  \langle \langle \widetilde{\Phi}_{\alpha}(t)| \hat{S}_{t} \hat{J}_{z} \hat{S}^{-1}_{t}  |\widetilde{\Phi}_{\alpha}(t)\rangle \rangle \nonumber \\
&= - \frac{2}{N} \sum_{m,n} |C_{-m,n}^{\alpha}|^{2} m = -\overline{J_{z}^{\alpha} } = 0,
\end{align}
where $|\widetilde{\Phi}_{\alpha}(t)\rangle \rangle = \hat{S}_{t} |\Phi_{\alpha}(t)\rangle \rangle$ and we have used the fact that $\hat{S}_{t} \hat{J}_{z} \hat{S}_{t}^{-1} = - \hat{J}_{z}$ along with Eqs.\eqref{cof_TR_FM} and \eqref{cof_TR_FM2}.

Hence, for the cases where the driving phase $\phi=\pi/2$ or $\phi = 3\pi/2 $, the API $\overline{J_{z} } = \sum_{\alpha} P_{\alpha} \overline{J_{z}^{\alpha} } = 0$ for any arbitrary initial condition. In order to achieve a non-zero API, we must therefore not only have $E_{2} \neq 0$ but also $\phi \neq \pi/2 $ and $\phi \neq 3\pi/2 $.

\subsubsection{Dependence of API on the driving phase} \label{API_two_phi}
Having investigated the symmetries of the Floquet Hamiltonian and ways to break them, let us now discuss how the value of $\overline{J_{z}^{\alpha} }$ depends on the driving phase $\phi$. At first, we note that two Floquet Hamiltonians which are related by a symmetry transformation have the same quasi-energy (QE) spectrum. We consider two Floquet Hamiltonians $\hat{H}_{F1}(t)$ and $\hat{H}_{F2}(t)$ satisfying
\begin{align}
\hat{H}_{F1}(t) | \Phi_{\alpha}^{(1)}(t)  \rangle &= \epsilon_{\alpha}^{(1)} |\Phi_{\alpha}^{(1)}(t) \rangle, \nonumber \\
\hat{H}_{F2}(t) | \Phi_{\alpha}^{(2)}(t)  \rangle &= \epsilon_{\alpha}^{(2)} |\Phi_{\alpha}^{(2)}(t) \rangle, 
\end{align}
with $|\Phi_{\alpha}^{(i)}(t) \rangle \rangle$ and $\epsilon_{\alpha}^{(i)}$ ($i = 1,2$) being the associated FMs and QEs. We assume that $\hat{H}_{F1}(t)$ and $\hat{H}_{F2}(t)$ are connected via a symmetry transformation $\hat{H}_{F1}(t) = \hat{S}\hat{H}_{F2}(t) \hat{S}^{-1}$, with $\hat{S}$ being the corresponding symmetry operator. This gives rise to
\begin{align}
\hat{S}\hat{H}_{F2}(t) \hat{S}^{-1}  \hat{S} | \Phi_{\alpha}^{(2)}(t) \rangle  &= \hat{H}_{F1}(t) \hat{S} | \Phi_{\alpha}^{(2)}(t) \rangle  \nonumber \\
&= \epsilon_{\alpha}^{(2)} \hat{S} |\Phi_{\alpha}^{(2)}(t) \rangle . \label{FMs_two}
\end{align}
which implies that $\epsilon_{\alpha}^{(2)}$ and $\hat{S} |\Phi_{\alpha}^{(2)}(t) \rangle $ are the eigenvalue and eigenstate for $\hat{H}_{F1}(t)$ as well. In this way, we demonstrate that $\hat{H}_{F1}(t)$ and $\hat{H}_{F2}(t)$ share the same QE spectrum. Moreover, for the non-degenerate Hamiltonians $\hat{H}_{F1}(t)$ and $\hat{H}_{F2}(t)$, this further implies that $\hat{S} | \Phi_{\alpha}^{(2)}(t) \rangle \rangle$ can only differ from $|\Phi_{\alpha}^{(1)}(t) \rangle \rangle$ by at most a phase factor.

For two different driving phases $\phi$ and $-\phi$, if we further consider $\hat{H}_{F1}(t) = \hat{H}_{F}(t, \phi)$ and $\hat{H}_{F2}(t) = \hat{H}_{F}(t, -\phi)$, these two Hamiltonians are related by the symmetry operator
\begin{equation}
\hat{S}_{\phi}^{I} = \hat{R}_{y}(\pi) \otimes \hat{\Theta}
\end{equation}
which yields the transformation
\begin{equation}
S_{\phi}^{I}:  (J_{x}, J_{z}, t)  \rightarrow (J_{x}, J_{z}, -t). \label{S_P_I}
\end{equation}
Hence, it immediately follows that \cite{QM_Sakurai}
\begin{align}
\overline{J_{z}^{\alpha} } (\phi)  &= \frac{2}{N} \langle \langle \Phi_{\alpha}^{(1)}(t) |\hat{J}_{z} | \Phi_{\alpha}^{(1)}(t) \rangle \rangle  \nonumber \\
&= \frac{2}{N} \langle \langle \widetilde{\Phi}_{\alpha}^{(1)}(t)  | \hat{S}_{\phi}^{I} \hat{J}_{z} (\hat{S}_{\phi}^{I} )^{-1} | \widetilde{\Phi}_{\alpha}^{(1)}(t) \rangle \rangle \nonumber \\
& = \frac{2}{N} \langle \langle \Phi_{\alpha}^{(2)}(t) | \hat{J}_{z} | \Phi_{\alpha}^{(2)}(t)\rangle \rangle   = \overline{J_{z}^{\alpha} } (- \phi),
\end{align}
where $| \widetilde{\Phi}_{\alpha}^{(1)}(t) \rangle \rangle =  \hat{S}_{\phi}^{I} | \Phi_{\alpha}^{(1)}(t) \rangle \rangle  = \eta | \Phi_{\alpha}^{(2)}(t) \rangle \rangle$ with $\eta$ being an arbitrary phase factor and we have employed the relation $\hat{S}_{\phi}^{I} \hat{J}_{z} (\hat{S}_{\phi}^{I} )^{-1} = \hat{J}_{z} $. This shows that the contribution $\overline{J_{z}^{\alpha} } (\phi)$ to the API from each FM  possess a \textit{mirror symmetry} around $\phi = 0$ and $\phi = \pi$, where we have noticed that $\overline{J_{z}^{\alpha} } (\phi)$ is periodic in $\phi$ with period $2\pi$. It is also important to emphasize once again that the above conclusion relies on the assumption that $\hat{H}_{F}(t)$ is non-degenerate, which, as previously mentioned, is well-justified by the extension of the von Neumann-Wigner theorem  \cite{ Ratchet_rev, Non_degenerate}. 

Similarly, if we consider $\hat{H}_{F1}(t) = \hat{H}_{F}(t, \phi)$ and $\hat{H}_{F2}(t) = \hat{H}_{F}(t, \phi + \pi)$, the symmetry operation 
\begin{equation}
S_{\phi}^{II}:  (J_{x}, J_{z}, t)  \rightarrow (J_{x}, -J_{z}, t + T/2),
\end{equation}
transforms $\hat{H}_{F1}(t)$ into $\hat{H}_{F2}(t)$, with the symmetry operator being
\begin{equation}
\hat{S}_{\phi}^{II}= \hat{R}_{x}(\pi) \otimes \hat{Q}(T/2).
\end{equation}
Since $\hat{S}_{\phi}^{II}$ reflects $J_{z}$ as $\hat{S}_{\phi}^{II} \hat{J}_{z} (\hat{S}_{\phi}^{II} )^{-1} = -\hat{J}_{z} $, it results in $\overline{J_{z}^{\alpha} } (\phi) = - \overline{J_{z}^{\alpha} } (\phi + \pi)$. Hence in addition to the mirror symmetry, $\overline{J_{z}^{\alpha} } (\phi)$ also possess a \textit{shift anti-symmetry}.

\subsection{Classical limit} \label{classical_limit}
In the limit of infinite particle number $N \rightarrow \infty$ and small interaction energy $U_{BH} \rightarrow 0$, such that $\Lambda = N U_{BH}$ is fixed, the dynamics of the particles can be well described by that of a classical non-rigid pendulum \cite{BJJ_Rabi_1, BJJ_Rabi_2, BJJ_Rabi_3, BJJ_Driven_1}. In order to explore the behavior of the API in this classical limit, we adopt the mean-field approximation as $\hat{a}_{j} = a_{j}$ $( j = L,R)$, with $a_{j} $ being a $c$-number \cite{GPE_1}. Since the total particle number $N_{L}^{2} + N_{R}^{2}  = N$ is conserved, it is convenient to express $a_{j}$ in the phase-density representation $a_{j} = \sqrt{N_{j}} e^{i \theta_{j}}$, where the particle numbers $N_{j}$ and the phases $\theta_{j}$ are in general time-dependent. We further introduce the two conjugate variables 
\begin{align}
Z(t) &= (N_{L} - N_{R}) /N,  &Z \in [-1,1]  \nonumber \\
\varphi(t) &= \theta_{R} - \theta_{L} \label{cl_z_phi},
\end{align}
representing the relative population imbalance between the two wells and the relative phase difference, respectively. Substituting $Z$ and $\varphi$ into the Eq.\eqref{BH_model} and replacing all the operators $\hat{a}_{j}$ ($\hat{a}_{j}^{\dagger}$) by $a_{j}$ ($a_{j}^{*}$), we obtain the classical Hamiltonian 
\begin{equation}
H_{cl}(t) = \frac{\Lambda}{2} Z^{2} - \sqrt{1-Z^{2}} \text{cos}\varphi + 2f(t) Z,  \label{BH_classical}
\end{equation}
which describes a driven non-rigid pendulum with angular momentum $Z$ and length proportional to $\sqrt{1-Z^{2}} $ \cite{BJJ_Rabi_1, BJJ_Rabi_2, BJJ_Rabi_3, BJJ_Driven_1}. $\Lambda = NU_{BH}$ is the coupling strength which is inversely proportional to the effective mass of the pendulum. The corresponding equations of motion are thus
\begin{align}
\dot{Z} &= - \sqrt{1-Z^{2}} ~\text{sin}\varphi, \nonumber \\
\dot{\varphi} &= \Lambda Z +  \frac{Z}{\sqrt{1-Z^{2}}}~\text{cos}\varphi + 2f(t). \label{eom_classical}
\end{align}
Such a classical reformulation allows us to interpret the API as the average angular momentum of the pendulum 
\begin{equation}
\overline{\Delta \rho} = \overline{Z} = lim_{\tau,  \tau^{\prime} \rightarrow \infty} ~\frac{1}{ \tau^{\prime} } \int_{\tau}^{\tau + \tau^{\prime}} Z(t) dt. \label{API_cl}
\end{equation}

\begin{figure}
  \centering
  \includegraphics[width=0.45\textwidth]{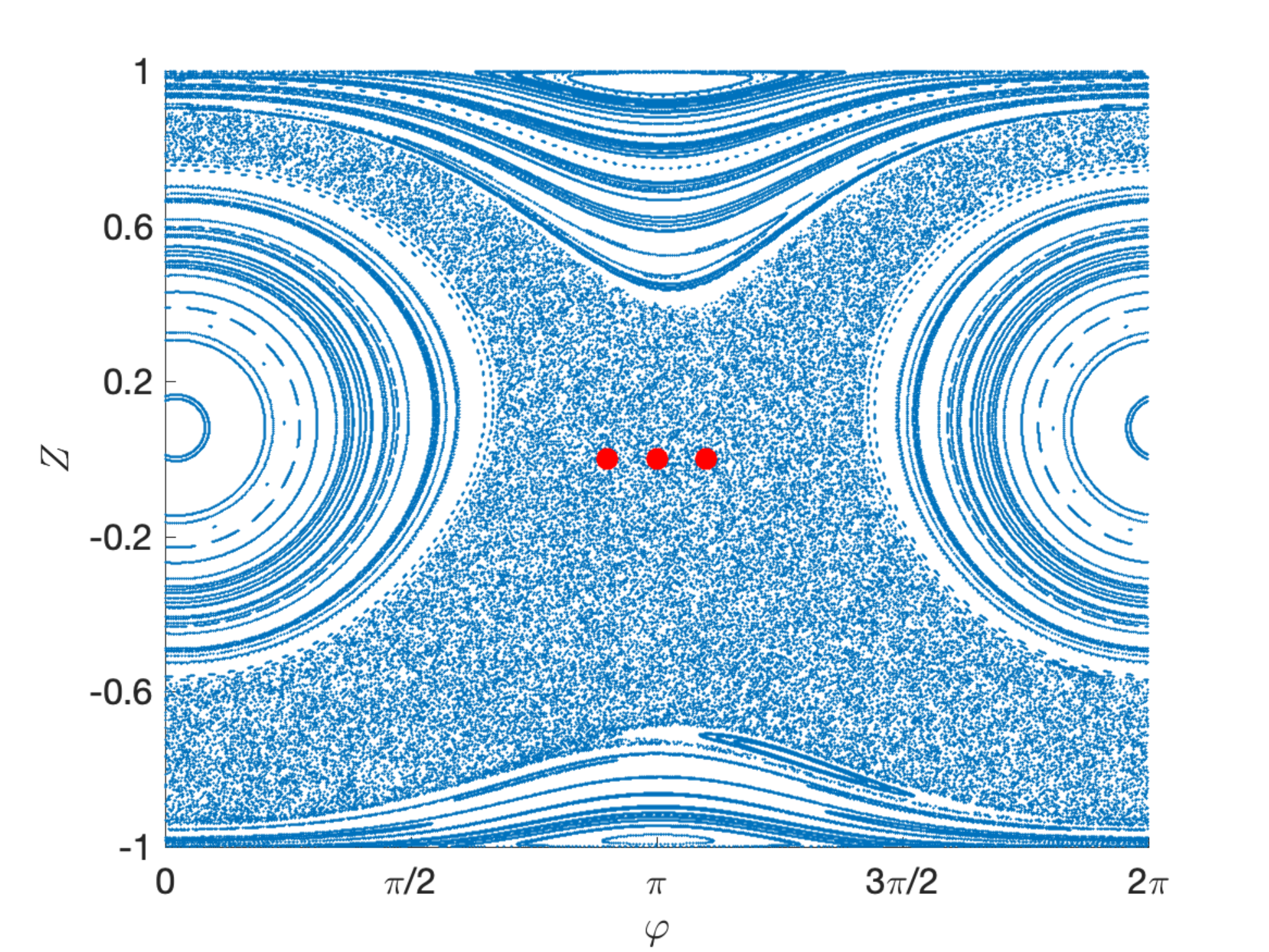}\hfill
  \caption{(Color online) Poincar\'e surfaces of sections (PSOS) for $\Lambda = 5$, the parameters for the driving force $f(t)$ are $E_{1} = 0.4$, $E_{2} = 0.2$, $\omega = 0.5$ and $\phi = 0$. The three red dots from left to right denote the phase space point $(Z = 0, \varphi = 9\pi/10)$, $(Z = 0, \varphi = \pi)$, $(Z = 0, \varphi = 11\pi/10)$, respectively, which will be used as the initial conditions for the classical simulations (see discussions below).}
\label{ps_cl}
\end{figure}

The particle dynamics in the classical limit can be well understood through an analysis of the three dimensional (3D) phase space characterized by $(Z,\varphi,t)$ underlying the equations of motion Eq.\eqref{eom_classical}. The stroboscopic Poincar\'e surfaces of sections (PSOS) (see Fig.~\ref{ps_cl}) of the particle dynamics reveal that the system has a mixed phase space that depends on the choice of the system parameters with both chaotic and regular components separated by Kolmogorov-Arnold-Moser (KAM) tori \cite{Ratchet_rev, PS}. Due to ergodicity, a trajectory initialized anywhere in the chaotic layer explores the entire chaotic layer in the course of its dynamics. The average $\overline{Z}$ for such a trajectory, corresponding to the value of API for the chosen initial condition, can thus be non-zero only if we break all the symmetries of the equations of motion Eq.\eqref{eom_classical} that transforms $Z \rightarrow -Z$.

From Eq.\eqref{eom_classical}, it can be seen that the system is invariant with respect to the generalized parity transformation 
\begin{equation}
S_{p}: (Z, \varphi, t) \rightarrow (-Z, -\varphi, t+\tau)
\end{equation}
if the driving law has the symmetry $f(t) = -f(t + \tau)$ for any arbitrary time shift $\tau$. On the other hand, if the driving law satisfy $f(t) = -f(-t + \tau)$, the generalized parity and time reversal operation
\begin{equation}
S_{pt}: (Z, \varphi, t) \rightarrow (-Z, \varphi, -t+\tau)
\end{equation}
keep the system invariant. Since these are the only two possible symmetry transformations of the system which flips the sign of $Z$, one needs to break them in order to achieve a non-zero API. A bi-harmonic driving force with $E_{1}, E_{2} \neq 0$ and $\phi \neq n\pi/2$ ( $n \in$  odd integers) breaks both symmetries $S_{p}$ and $S_{pt}$ thus allowing for a non-vanishing API. Furthermore, we can also predict the dependence of the API $\overline{Z}$ on the driving phase $\phi$ by a similar symmetry analysis. We note that the Eq.\eqref{eom_classical} is invariant under the joint transformation
\begin{align}
&\phi \rightarrow -\phi,  &(Z, \varphi, t) \rightarrow (Z, -\varphi, -t) \label{mirror_z}
\end{align}
Hence it follows that $\overline{Z}$ should possess a mirror symmetry with respect to $\phi$, i.e. $\overline{Z}(\phi)  = \overline{Z}( - \phi)$. The joint transformation
\begin{align}
&\phi \rightarrow \phi+ \pi,  &(Z, \varphi, t) \rightarrow (-Z, -\varphi, t+T/2) \label{shift_z}
\end{align}
also keeps the equation of motion invariant, hence $\overline{Z}$ has a shift anti-symmetry $\overline{Z}(\phi)  = -\overline{Z}(\phi + \pi)$.

Before closing this section, we note that the above mean-field approximation $\hat{a}_{j} = a_{j} = \sqrt{N_{j}} e^{i \theta_{j}}$ is equivalent to express the many-body wavefunction as \cite{GPE_1}
\begin{equation}
\Psi(x_1, x_2, ..., x_N,t) = \prod_{i=1}^{N} \phi({x_{i},t)}, \label{GP_wf}
\end{equation}
with the single-particle state
\begin{equation}
\phi({x_{i},t)} = c_{L}(t) u_{L}(x_{i}) + c_{R}(t) u_{R}(x_{i}), \label{GP_single}
\end{equation}
being the linear superposition of the localized states $u_{L}(x)$ and $u_{R}(x)$. The time-dependent coefficients $c_{L/R}(t) $ are in general complex, fulfilling the normalization condition $|c_{L}(t)|^{2} + |c_{R}(t)|^{2} = 1$. The conjugate variables $Z(t)$ and $\varphi(t)$ can thus be expressed in terms of $ c_{L}(t)$ and $ c_{R}(t)$ as
\begin{align}
Z(t) &= |c_{L}(t)|^{2} - |c_{R}(t)|^{2},    \nonumber \\
\varphi(t) &= \text{arg}(c_{L}(t))- \text{arg}(c_{R}(t)) \label{cl_z_phi_wannier}.
\end{align}
This provides a relation between the dynamics for $Z(t)$ and $\varphi(t)$ and that of $ c_{L}(t)$ and $ c_{R}(t)$ respectively.

\section{Results} \label{Results_analysis}

\subsection{Initial state and numerical setup}
The initial condition in the classical limit is provided by a specific point ($Z$, $\varphi$) in the phase space, which determines the initial population and phase difference. In order to find its equivalent counterpart for the quantum limit, we employ the relations in Eqs.~\eqref{GP_wf} and \eqref{GP_single}, and express the many-body state as 
\begin{align}
|\theta, \varphi \rangle &= \frac{1}{\sqrt{N!}}\left[\text{cos}(\frac{\theta}{2}) \hat{a}^{\dagger}_{L} + \text{sin}(\frac{\theta}{2}) e^{i \varphi}\hat{a}^{\dagger}_{R}\right]^{N} ~|vac \rangle \nonumber \\
&= \sum_{N_L=0} ^{N} \left(\begin{array}{c} N \\ N_L \end{array} \right)^{1/2} \text{cos}^{N_L}(\theta/2)~\text{sin}^{N_R}(\theta/2) ~ e^{i N_R \varphi} ~|N_L,N_R \rangle,
\end{align}
which is the linear superposition of all the number states $\{|N_L, N_R \rangle\}$. The state $|\theta, \varphi \rangle$ is referred to as the atomic coherent state (ACS) \cite{ACS_1,ACS_2} fulfilling the completeness relation
\begin{equation}
(N+1) \int \frac{d \Omega}{4 \pi} |\theta, \varphi \rangle \langle \theta, \varphi | = 1, \label{completeness_ACS}
\end{equation}
with $d \Omega = \text{sin}\theta d \theta d \varphi$ being volume element. The ACS relates to the mean-field wavefunction $\Psi$ [c.f. Eqs. \eqref{GP_wf} and \eqref{GP_single}] as 
\begin{align}
\theta &= \text{cos}^{-1}(|c_{L}|^{2} - |c_{R}|^{2})  \nonumber\\
\varphi  &= \text{arg}(c_{L})- \text{arg}(c_{R})  \label{qm_z_phi_wannier}
\end{align}
where $\theta$ and $\varphi$ control the initial population difference $\text{cos}\theta = (N_{L} - N_{R})/N $ and the initial phase difference respectively. Comparing Eq. \eqref{qm_z_phi_wannier} to the Eq. \eqref{cl_z_phi_wannier}, we find a one-to-one correspondence between the $|\theta, \varphi \rangle$ and the ($Z$, $\varphi$) and thus allows us to compare the quantum and the classical dynamics. Correspondingly, the ACS can be expressed as
\begin{equation}
|\theta, \varphi \rangle = \sum_{m=-l} ^{l} \left(\begin{array}{c} 2l \\ m+l \end{array} \right)^{1/2} \text{cos}^{l+m}(\theta/2)~\text{sin}^{l-m}(\theta/2) ~ e^{i (l-m) \varphi} ~|l,m \rangle. \label{ACS_spin}
\end{equation}
in the angular momentum basis. In recent ultracold experiments, such an ACS can be implemented in a controllable manner. Tuning a two-photon transition between two hyperfine states of ${}^{87}\textrm{Rb}$ atoms allow us to prepare an ACS with arbitrary $|\theta, \varphi \rangle$ \cite{ACS_3,ACS_4}.

In this work, we aim to explore how the asymptotic population imbalance behaves when we go from the few-particle regime to the many-particle regime. To this end, we fix the coupling strength $\Lambda = NU_{BH} = 5.0$ for all our simulations and vary the interaction energy $U_{BH}$ and particle number $N$ accordingly. For all our quantum simulations, we choose the initial ACS $|\theta = \pi/2, \varphi \rangle$, which corresponds to $Z(0) = 0$ in the classical limit signifying a balanced particle population between the two wells at the beginning. The phase difference $\varphi$ is carefully chosen such that the ACS $|\theta, \varphi \rangle$ is always located within the chaotic layer corresponding to the classical PSOS [see three red dots in Fig.~\ref{ps_cl}]. We also simulate the classical limit by numerically integrating Eq.\eqref{eom_classical}. Finally, we compare the behavior of the API obtained from the quantum ($\overline{J}_{z}$) and classical ($\overline{Z}$) simulations.

\subsection{Variation of API with particle number and driving phase}
\begin{figure*}
  \centering
  \includegraphics[width=1.0\textwidth]{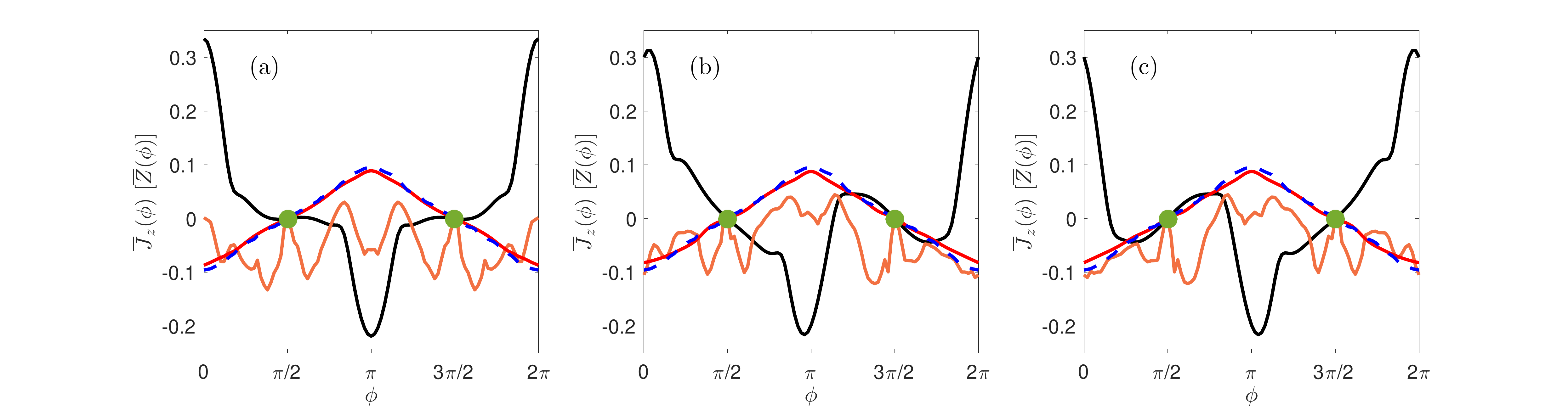}\hfill
     \caption{(Color online) Asymptotic population imbalance (API) as a function of the driving phase $\phi$ for the three initial states: (a) $|\Psi(0) \rangle =  |\pi/2, \pi \rangle$, (b) $|\Psi(0) \rangle =  |\pi/2, 9\pi/10 \rangle$, (c) $|\Psi(0) \rangle =  |\pi/2, 11\pi/10 \rangle$. The solid black, orange and red lines corresponds to particle number $N=2$, $N=20$ and $N=500$ respectively. The API in the classical limit is depicted as blue dashed lines for the corresponding initial conditions: (a) $ (Z=0, \varphi = \pi)$,  (b) $(Z=0, \varphi = 9\pi/10)$ and (c) $(Z=0, \varphi = 11\pi/10)$ in the classical PSOS (see Fig.~\ref{ps_cl}). The green solid dots indicate that the API vanishes at $\phi = \pi/2 $ and $\phi = 3\pi/2 $. Remaining parameters are $E_{1} = 0.4$, $E_{2} = 0.2$, $\omega = 0.5$, $\Lambda=5$. }
\label{APIs}
\end{figure*}

In Fig.~\ref{APIs}, we present the asymptotic population imbalance $\overline{J}_{z}$ as a function of the driving phase $\phi$ for different particle numbers $N = 2, 20, 500$ and different initial states $|\Psi(0) \rangle = |\pi/2, \pi \rangle$ [Fig.~\ref{APIs} (a)], $|\Psi(0) \rangle = |\pi/2, 9\pi/10 \rangle$ [Fig.~\ref{APIs} (b)] and $|\Psi(0) \rangle = |\pi/2, 11 \pi/10 \rangle$ [Fig.~\ref{APIs} (c)]. The API $\overline{Z}$ corresponding to the classical simulations for the same initial conditions [see three red dots in Fig.~\ref{ps_cl}] are depicted as well [all blue dashed lines in Fig.~\ref{APIs}]. We first discuss the results obtained for the classical limit. Since a trajectory initialized anywhere in the chaotic layer will explore the entire chaotic layer in the course of the dynamics due to ergodicity, it is hence guaranteed that the obtained value of API should be independent of the initial conditions. Hence, the observed behavior of $\overline{Z}$ is the same for all the three different initial conditions. As varying the driving phase $\phi$, $\overline{Z}(\phi)$ shows an oscillatory behavior having maxima (minima) at $\phi=\pi$ ($\phi=0, 2\pi$) and vanishes at $\phi=n\pi/2$ for all odd integers $n$. Most importantly,  it preserves both the mirror symmetry $\overline{Z}(\phi)  = \overline{Z}( - \phi)$ [see Eq.\eqref{mirror_z}] and the shift anti-symmetry $\overline{Z}(\phi)  = -\overline{Z}(\phi +\pi)$ [see Eq.\eqref{shift_z}], thus verifying our symmetry analysis in Sec. \ref{classical_limit}.

In the quantum limit, the behavior of the API $\overline{J}_{z}$ is much more complicated. For a large number of particles $N =500$, the behavior of $\overline{J}_{z}$ upon varying $\phi$ almost agrees very well with that of the API $\overline{Z}$ in the classical limit, independent of the initial quantum state [see the red solid lines in Fig.~\ref{APIs}]. As a result, $\overline{J}_{z}$ exhibits the corresponding mirror symmetry $\overline{J}_{z}(\phi)  = \overline{J}_{z}( - \phi)$ and shift anti-symmetry $\overline{J}_{z}(\phi)  = -\overline{J}_{z}(\phi +\pi)$ as well.

By contrast, the API in the few-particle regime depends strongly on the initial states. Most importantly, the symmetries of $\overline{J}_{z}(\phi)$ observed in the large particle limit are broken. For the initial state $|\Psi(0) \rangle = |\pi/2, \pi \rangle$, only the mirror symmetry is preserved [see, e.g., the black solid and the orange solid lines in Fig.~\ref{APIs} (a)], while for $|\Psi(0) \rangle = |\pi/2, 9\pi/10 \rangle$ or  $|\pi/2, 11\pi/10 \rangle$ both the mirror symmetry and the shift anti-symmetry are explicitly broken [c.f. Fig.~\ref{APIs} (b,c)]. Instead, a new symmetry which relates the value of $\overline{J}_{z}(\phi)$ for two different initial states is now observed in the few-particle regime. Specifically, the dependence of $\overline{J}_{z}$ on $\phi$ for the initial state $|\pi/2, 9\pi/10 \rangle$ [c.f. Fig.~\ref{APIs} (b)] can be obtained by a reflection of $\overline{J}_{z}(\phi)$ for the initial state $|\pi/2, 11\pi/10 \rangle$ [c.f. Fig.~\ref{APIs} (c)] about either $\phi=0$ or $\phi=\pi$. Since $|\pi/2, \varphi \rangle  = |\pi/2, \varphi-2\pi \rangle $, we can represent this symmetry by
\begin{equation}
[\overline{J}_{z}(\phi)]_{\varphi} = [\overline{J}_{z}(-\phi)]_{-\varphi}, \label{symmetry_two_initial_states}
\end{equation}
where $[\overline{J}_{z}(\phi)]_{\varphi}$ ($[\overline{J}_{z}(\phi)]_{-\varphi}$) denotes the obtained $\overline{J}_{z}$ value for the initial state $|\theta, \varphi \rangle$ ($|\theta, -\varphi \rangle$) for a given driving phase $\phi$. Lastly, we note that the API values vanish for $\phi = n\pi/2$ for all odd integers $n$ [see the green dots in Fig.~\ref{APIs}] in both the classical and the quantum limit in accordance with our symmetry analysis in Sec. \ref{St_symmetry} and Sec. \ref{classical_limit}.

\section{Discussions}\label{discussion}

\subsection{API in the few-particle regime} \label{API_few}
In order to explain the broken symmetries as well as the emergence of the new symmetry [see Eq.\eqref{symmetry_two_initial_states}] as we observed in the few-particle regime, we analyze the contribution of each Floquet mode to the value of the API. Specifically, since $\overline{J}_{z} (\phi) = \sum_{\alpha} P_{\alpha} (\phi) \overline{J_{z}^{\alpha} } (\phi) $ [c.f. Eq.\eqref{API_sum}], we inspect how each $P_{\alpha} $ and $\overline{J_{z}^{\alpha} } $ depend on the driving phase $\phi$. We note that while $\overline{J_{z}^{\alpha} } (\phi) $ is solely determined by the Floquet Hamiltonian, $P_{\alpha} (\phi) $ depends on both the Floquet Hamiltonian and the initial state. 

\begin{figure}
  \centering
  \includegraphics[width=0.5\textwidth]{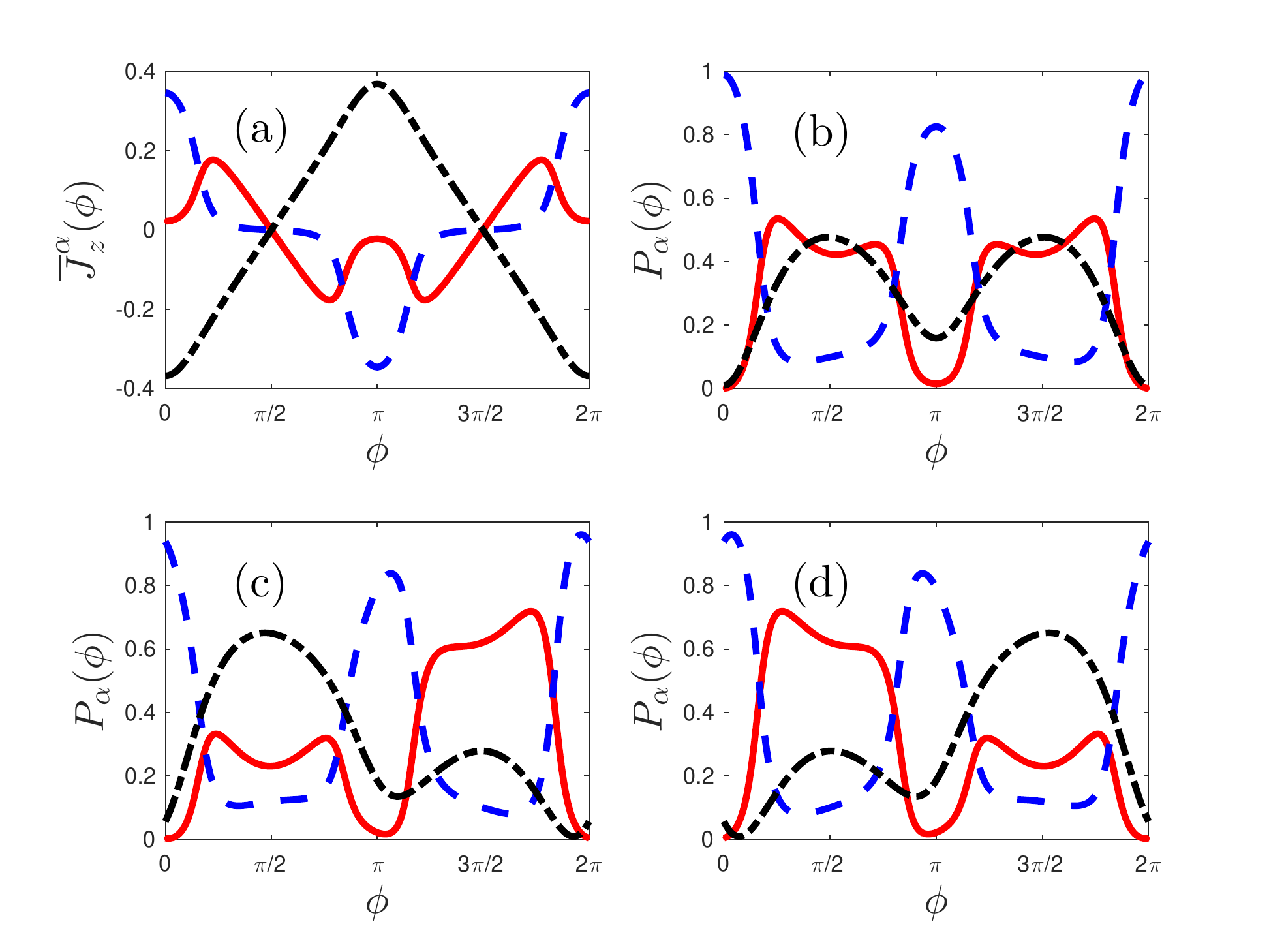}\hfill
  \caption{(Color online) Decomposition of $\overline{J}_{z} (\phi)$ with respect to three FMs for the case $N = 2$, in which (a) represents the $\overline{J_{z}^{\alpha} } (\phi)$,  (b-c) denote the $P_{\alpha}(\phi)$ for $|\Psi(0) \rangle =  |\pi/2, \pi \rangle$, $|\pi/2, 9\pi/10 \rangle$ and $|\pi/2, 11 \pi/10 \rangle$, respectively. The blue dashed, red solid and black dash-dotted line correspond to the FM $| \Phi_{1}(t)\rangle \rangle $, $| \Phi_{2}(t)\rangle \rangle $ and $| \Phi_{3}(t)\rangle \rangle $, respectively. The corresponding driving parameters are $E_{1} = 0.4$, $E_{2} = 0.2$ and $\omega = 0.5$.}
\label{floquet_N2}
\end{figure}

To illustrate this, we consider the case for $N=2$. Fig.~\ref{floquet_N2}(a) shows how the contributions $\overline{J_{z}^{\alpha} } $ from the three FMs depend on the driving phase $\phi$. As it can be seen, their dependence on $\phi$ are significantly different from each other, however all of them vanish for $\phi=n\pi/2$ for all odd integers $n$. Additionally, all the three $\overline{J_{z}^{\alpha} } (\phi) $ preserves both the mirror-symmetry and shift anti-symmetry, i.e., $\overline{J_{z}^{\alpha} } (\phi) =  \overline{J_{z}^{\alpha} } (-\phi)$ and $\overline{J_{z}^{\alpha} } (\phi) = - \overline{J_{z}^{\alpha} } (\phi + \pi)$. Hence, the broken symmetries of $\overline{J}_{z}$ in the few particle regime are definitely not due to the contributions from $\overline{J_{z}^{\alpha} } (\phi)$ as already verified by our previous symmetry analysis but stem from the weights $P_{\alpha} (\phi) $. In Fig.~\ref{floquet_N2}(b-d), we show the behavior of $P_{\alpha} (\phi) $ corresponding to the three initial states $|\pi/2, \pi \rangle$, $ |\pi/2, 9\pi/10 \rangle$ and $ |\pi/2, 11\pi/10 \rangle$, respectively. Indeed, as one can see the exhibited symmetric (asymmetrical) structure for $P_{\alpha}(\phi)$ results in the mirror symmetry (symmetry-breaking) in the corresponding $\overline{J}_{z}(\phi)$. For instance, $P_{\alpha}(\phi) = P_{\alpha}(-\phi) $ for initial state $|\pi/2, \pi \rangle$, hence $\overline{J}_{z}$ fulfills $\overline{J}_{z}(\phi) = \overline{J}_{z}(-\phi)$. By contrast,  $P_{\alpha}(\phi) \neq P_{\alpha}(-\phi) $ for the initial states $|\pi/2, \pi \pm \pi/10 \rangle$, which results in $\overline{J}_{z}(\phi) \neq \overline{J}_{z}(-\phi)$. Moreover, since $P_{\alpha}(\phi)$ does not obey the property $P_{\alpha}(\phi) = P_{\alpha}(\phi + \pi)$ in general, it thereby explains the broken shift anti-symmetry for all the $\overline{J}_{z}(\phi)$ in the few-particle regime. 

The emergence of the new symmetry in Eq.\eqref{symmetry_two_initial_states} can also be understood from the behavior of $P_{\alpha}(\phi)$. Since for two different initial states $|\theta, \varphi \rangle$ and $|\theta, -\varphi \rangle$, the corresponding $P_{\alpha}(\phi)$ satisfy $[P_{\alpha}(\phi)]_{\varphi}  = [P_{\alpha}(-\phi)]_{-\varphi}$ [see Fig.~\ref{floquet_N2}(c,d) and Appendix \ref{Appendix_2}], hence
\begin{align}
[\overline{J}_{z}(\phi)]_{\varphi} &= \sum_{\alpha} [P_{\alpha}(\phi)]_{\varphi}  \overline{J_{z}^{\alpha} } (\phi) \nonumber \\
&= \sum_{\alpha} [P_{\alpha}(-\phi)]_{-\varphi}  \overline{J_{z}^{\alpha} } (-\phi) =  [\overline{J}_{z}(-\phi)]_{-\varphi}. \label{symmetry_Jz_two_initial}
\end{align}
Here, we have employed the mirror symmetry property of $ \overline{J_{z}^{\alpha} } $, along with the fact that $ \overline{J_{z}^{\alpha} } $ is independent for different choices of the initial states.

\subsection{API in the many-body regime}
We now discuss the behavior of the API in the many-particle regime in detail. Although the dependence of the API $\overline{J}_{z}$ on the driving phase $\phi$ for $N=500$ agrees very well with that of the $\overline{Z}$ in the classical limit (see Fig.~\ref{APIs}), we show now that there exists a significant disagreement in the corresponding real-time population imbalance due to quantum correlations.

\subsubsection{Quantum correlations}
\begin{figure}
  \centering
  \includegraphics[width=0.5\textwidth]{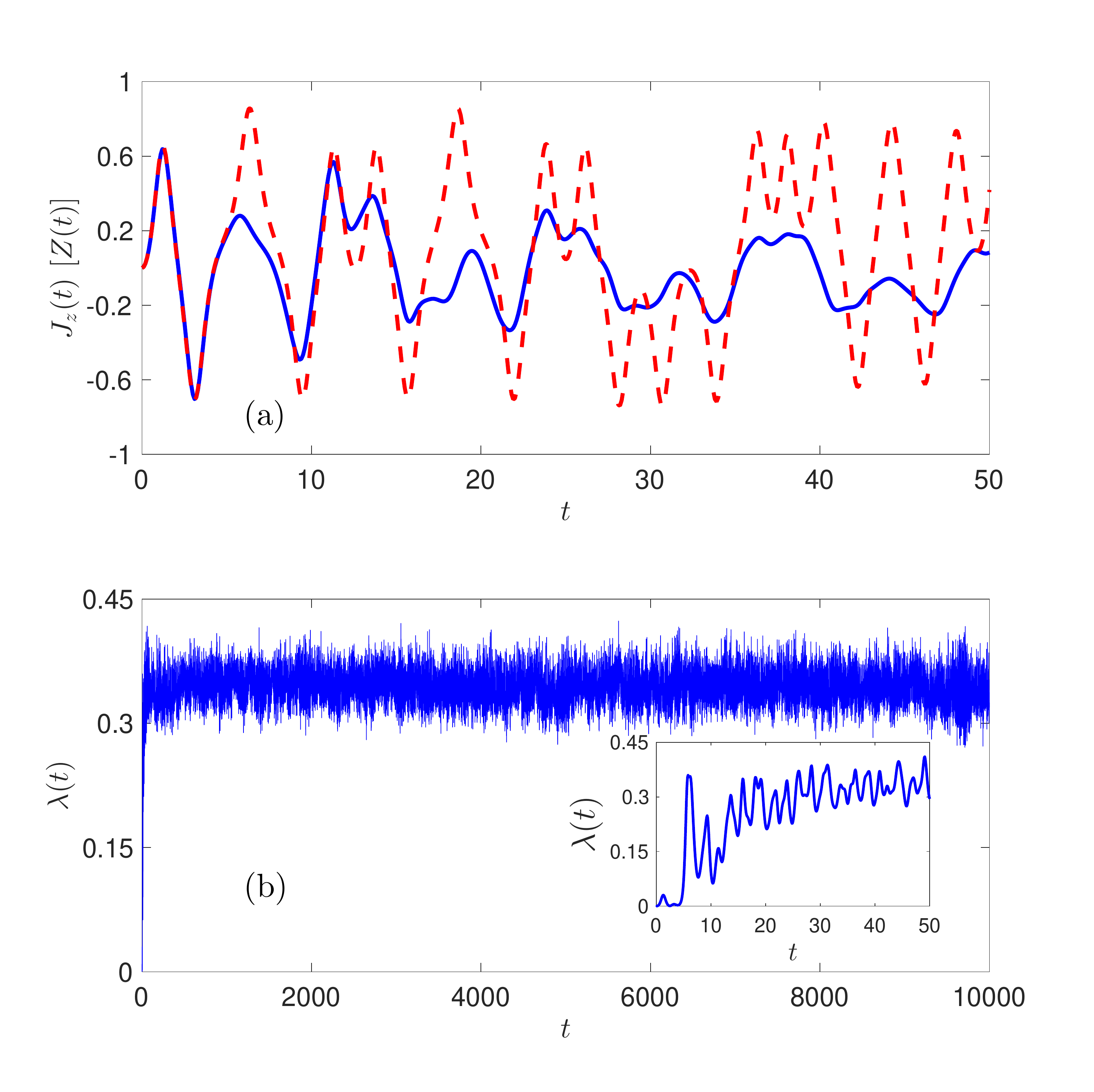}\hfill
  \caption{(Color online) Upper panel: Real-time dynamics for $J_{z}(t)$ (blue solid line) and $Z(t)$ (red dashed line) for the initial condition $|\Psi(0) \rangle =  |\pi/2, \pi \rangle$ for the quantum limit and $(Z(0)=0, \varphi(0) = \pi)$ for the classical limit. Lower panel: quantum depletion $\lambda(t)$ for the case examined in the upper panel, and the inset denotes the transient dynamics for $\lambda(t)$ for $t<50$. Both of them correspond to the particle number $N=500$ and the values of the driving parameters are $E_{1} = 0.4$, $E_{2} = 0.2$, $\omega = 0.5$ and $\phi = 0$. }
\label{gpop_mx_cl}
\end{figure}

In Fig.~\ref{gpop_mx_cl} (a), we show the time evolution of $J_{z}(t)$ corresponding to $N=500$ for the initial state $|\Psi(0) \rangle = |\pi/2, \pi \rangle$ along with that of $Z(t)$ for the initial condition $(Z(0)=0, \varphi(0) = \pi)$. Note that for $N=500$, the system is already in the weak-interaction regime, with $J_{BH}/U_{BH} = 50 \gg 1$, which, as one may anticipate, renders the mean-field approximation to work well \cite{GPE_1,GPE_2}. We observe that although the two quantities agree very well for very short timescales ($t \leq 5$), they evolve much differently at longer timescales. In order to understand why such a deviation occurs, we perform a spectral decomposition of the reduced one-body density operator \cite{dma1_1,dma1_2}
\begin{equation}
\hat{\rho}_{1}(t) = \sum_{i=1}^{2} n_{i}(t) | \phi_{i} (t)\rangle \langle \phi_{i}(t)|,
\end{equation}
and monitor the evolution for the quantum depletion defined as $\lambda(t) = 1 - n_{1}(t)$. Here $\{n_{i}(t)\}$ are the normalized time-dependent natural populations sorted in a descending order of their values such that $n_{1}(t) \geqslant n_{2}(t)$. $\{| \phi_{i} (t)\rangle\} $ denote the natural orbitals that form a time-dependent single-particle basis for the description of the dynamical system. Note that the two-mode expansion of the field operator $\hat{\psi}(x)$ in Eq.\eqref{2_mode_psi} leads to the single-particle Hamiltonian being restricted to a two-dimensional Hilbert space and thus gives rise to only two natural populations (natural orbitals) in the spectral decomposition. Physically, the natural population $n_{i}(t)$ denotes the probability for finding a single particle occupying the state $ | \phi_{i}(t) \rangle$ at time $t$, after tracing out all other $(N-1)$ particles. When $\lambda(t) = 0$, all the bosons reside in the single-particle state $\phi({x_{i},t)}$ [c.f. Eq.\eqref{GP_single}]. Hence the corresponding many-body wavefunction can be expressed in a mean-field product form [c.f. Eq.\eqref{GP_wf}]. According to our discussions in Sec. \ref{classical_limit}, this implies that the time evolution of the quantum dynamics $J_{z}(t)$ is completely equivalent to that of the classical dynamics $Z(t)$. In contrast for $\lambda(t) > 0$, quantum correlations come into play and therefore this would result in a completely different dynamics between $J_{z}(t)$ and $Z(t)$. This is indeed seen in the evolution of $\lambda(t)$ shown in Fig.~\ref{gpop_mx_cl} (b). For short timescales $t \leq 5$, $\lambda \approx 0$ as a result of which $J_{z}(t)$ and $Z(t)$ evolve in the same manner. However, for $t > 5$, the value of $\lambda (t)$ increases rapidly resulting in the different time evolution of  the $J_{z}(t)$ and the $Z(t)$ dynamics. Hence the existing quantum correlations in the system lead to significant quantitative differences between the quantum and classical dynamics although the time averaged asymptotic particle imbalance is the same in both cases.

\subsubsection{Time-averaged Husimi distribution}

\begin{figure}
  \centering
  \includegraphics[width=0.5\textwidth]{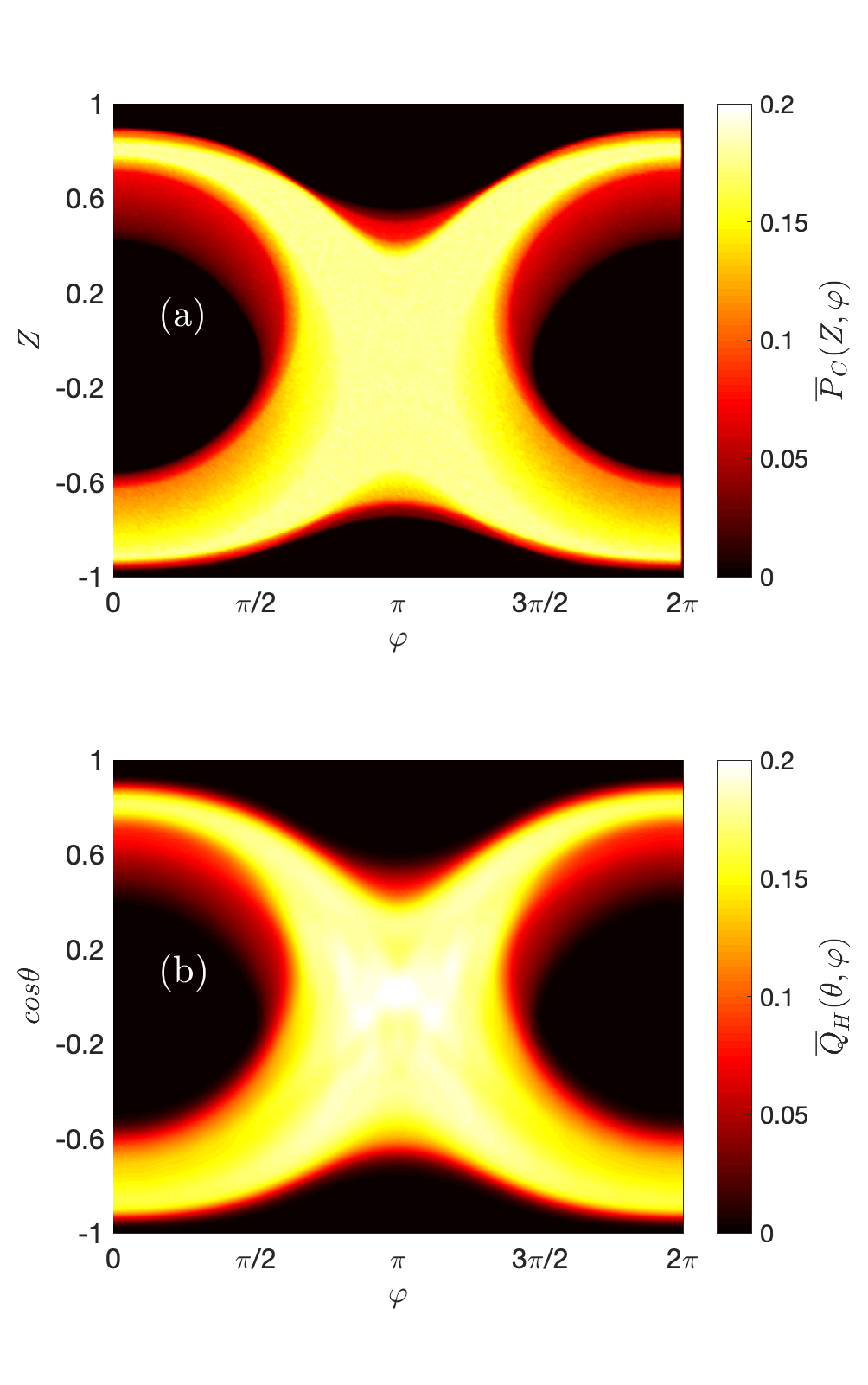}\hfill
  \caption{(Color online) Upper panel: Phase space probability distribution function (PDF) for a classical trajectory initialized at the point ($Z = 0, \varphi = \pi$). Lower panel: The time-averaged Husimi distribution for $N=500$ and for the initial state $|\Psi(0) \rangle = |\pi/2, \pi \rangle$. Note that the $\theta$-axis has been rescaled to $\text{cos}\theta$ in accordance with the correspondence $\text{cos}\theta = Z$. The values of the driving parameters are $E_{1} = 0.4$, $E_{2} = 0.2$, $\omega = 0.5$ and $\phi = 0$. }\label{husimi}
\end{figure}

This leads to the interesting but non-trivial question: while large discrepancies persist between the dynamics of $Z(t)$ and $J_{z}(t)$, how does it eventually result in the same value of the time averaged quantities $ \overline{Z}$ and $\overline{J}_{z}$? In order to answer this question, we first explore how does our classical state initialized at $(Z = 0, \varphi = \pi)$ evolve over time in the phase space up to $t=10^{7}$. Since the initial state belongs to the chaotic layer in the PSOS (Fig.~\ref{ps_cl}), it explores the entire chaotic sea ergodically in the course of its dynamics. In Fig.~\ref{husimi}(a), we show the probability density function (PDF) $\overline{P}_{C} (Z, \varphi)$ of this trajectory over the entire course of the dynamics. Note that the PDF $\overline{P}_{C} (Z, \varphi)$ unsurprisingly bears a striking resemblance with the corresponding PSOS in Fig.~\ref{ps_cl}. Since the system visits all the possible states (phase space points) that belong to the chaotic layer ergodically, it results in the uniform distribution of $\overline{P}_{C} (Z, \varphi)$ for all $(Z, \varphi)$ belonging to the chaotic sea. The regions for $\overline{P}_{C} (Z, \varphi) = 0$ correspond to the regular islands which the system can not enter. The visualization of the $Z(t)$ dynamics in terms of the PDF $\overline{P}_{C} (Z, \varphi)$ allows us to reformulate the time-averaged population imbalance $\overline{Z}$ as
\begin{equation}
\overline{Z} = lim_{\tau,  \tau^{\prime} \rightarrow \infty} ~\frac{1}{ \tau^{\prime} } \int_{\tau}^{\tau + \tau^{\prime}} Z(t) dt = \int d \sigma \overline{P}_{C} (Z, \varphi) Z, \label{API_PDF}
\end{equation}
where $d \sigma $ is volume element of the phase space. Averaged over the whole dynamics, $\overline{P}_{C} (Z, \varphi) d \sigma $ thus indicates the probability for the system to be located at the state $(Z, \varphi )$. 

In the quantum limit, the evolution of the initial state $|\Psi(0) \rangle = |\pi/2, \pi \rangle$ for $N=500$ can be visualized, analogous to $\overline{P}_{C} (Z, \varphi)$, by the time-averaged Husimi distribution (TAHD) defined as \cite{TAHD_1,TAHD_2}
\begin{equation}
\overline{Q}_{H}(\theta, \varphi)  = lim_{\tau,  \tau^{\prime} \rightarrow \infty} ~\frac{1}{ \tau^{\prime} } \int_{\tau}^{\tau +  \tau^{\prime}} Q_{H} (\theta, \varphi,t) dt,
\end{equation}
where  
\begin{equation}
Q_{H} (\theta, \varphi,t) = \frac{N+1}{4\pi}  \langle \theta, \varphi | \hat{\rho}(t) | \theta, \varphi \rangle, \label{husimi_t}
\end{equation}
with $ \hat{\rho}(t) = | \Psi(t) \rangle \langle \Psi(t)| $ being the system's density matrix. $Q_{H} (\theta, \varphi,t) $ thus satisfies the normalization condition $\int  Q_{H} (\theta, \varphi,t) d \Omega  = 1$. The TAHD represents the probability for our quantum system locating at the ACS $|\theta, \varphi\rangle$ averaged over the entire dynamics. As can be seen from  Fig.~\ref{husimi}(b), the TAHD matches very well with the distribution $\overline{P}_{C} (Z, \varphi)$ in Fig.~\ref{husimi}(a). Note that, the $\theta$-axis in Fig.~\ref{husimi}(b) has been rescaled to $\text{cos}\theta$ since $\text{cos}\theta = Z$ [c.f. Eq.\eqref{cl_z_phi_wannier} and Eq.\eqref{qm_z_phi_wannier}]. This suggests that the system evolves in an ergodic manner such that it has an equal probability for occupying all the ACSs located in the corresponding classical chaotic sea in the course of the dynamics. Analogous to the classical case [c.f. Eq.~\eqref{API_PDF}], the API in the quantum limit can be reformulated in terms of the TAHD $\overline{Q}_{H} (\theta, \varphi)$ as (see Appendix \ref{Appendix_3})
\begin{align}
\overline{J_{z} } &= lim_{ \tau,  \tau^{\prime} \rightarrow \infty} ~\frac{2}{N \tau^{\prime} } \int_{\tau}^{\tau +  \tau^{\prime}}dt~ \langle \hat{J}_{z}\rangle(t) \nonumber \\
&= \int d \Omega ~\overline{Q}_{H} (\theta, \varphi)  J_{z}(\theta, \varphi), \label{API_husimi}
\end{align}
with $J_{z}(\theta, \varphi) =  \frac{2}{N}\langle \theta, \varphi | \hat{J}_{z} | \theta, \varphi \rangle$. Since the TAHD $\overline{Q}_{H} (\theta, \varphi)$ has a similar distribution as the classical PDF $\overline{P}_{C} (Z, \varphi)$,  together with the fact $J_{z}(\theta, \varphi) = \text{cos}\theta = Z$, the value of API thus agrees in both the quantum and classical limit.

\section{Conclusions and Outlook} \label{Conclusions}
We have investigated a driven many-body bosonic ensemble confined in a 1D double-well potential and showed how an asymptotic population imbalance of particles between the two wells emerges from an initially symmetric particle population in both the quantum and classical limits. The asymptotic population imbalance can be controlled by changing the phase of the driving force as well as the total number of particles in the setup. The variation of the API in the few-particle quantum regime is elaborated in terms of the symmetries of the underlying Floquet modes. In the many-particle regime, the API can be interpreted in terms of an equivalent classical driven non-rigid pendulum. However, we show that quantum correlations still exist in the many-body system resulting in significant differences in the real-time evolution of the particle population imbalance as compared to the corresponding classical description. Possible future investigations include the study of API for an atomic mixture consisting of two atomic species with different mass and interactions. The effect from the higher bands for the double-well potential, beyond the single-band approximation discussed here, is also an interesting perspective.

\begin{acknowledgments}
The authors acknowledge fruitful discussions with Kevin Keiler. J.C. and P.S. gratefully acknowledge financial support by the Deutsche Forschungsgemeinschaft (DFG) in the framework of the SFB 925 ``Light induced dynamics and control of correlated quantum systems”. The excellence cluster ``The Hamburg Centre for Ultrafast Imaging-Structure: Dynamics and Control of Matter at the Atomic Scale” is acknowledged for financial support. A.K.M acknowledges a doctoral research grant (Funding ID: 57129429) by the Deutscher Akademischer Austauschdienst (DAAD).
\end{acknowledgments}

\appendix
\section{Relative orders for the operators in the $\hat{S}_{t}$ operator} \label{Appendix_1}
In this part, we demonstrate that the three operators among the symmetry operator $\hat{S}_{t}$ in Eq.\eqref{S_t} commute with each other, therefore, any changes of the relative orders among them do not affect the physics. Recall the form of the $\hat{S}_{t}$ operator $\hat{S}_{t} = \hat{R}_{z}(\pi) \otimes \hat{\Theta} \otimes \hat{Q}(T/2)$. It apparently shows that $\hat{R}_{z}(\pi)$ commutes with $\hat{Q}(T/2)$ since they are acting on different Hilbert spaces. Next, we illustrate that the rotation operator $ \hat{R}_{z}(\pi)$ commutes with the time-reversal operator $\hat{\Theta}$ as well. For an arbitrarily general state $|\psi \rangle$, it follows that 
\begin{equation}
\hat{\Theta}  \hat{R}_{z}(\pi) |\psi \rangle = \hat{\Theta} e^{-i \pi \hat{J}_{z}} |\psi \rangle = \left[\hat{\Theta} e^{-i \pi \hat{J}_{z}} \hat{\Theta}^{-1} \right]\hat{\Theta} |\psi \rangle,
\end{equation}
since $\hat{\Theta} \hat{J}_{z} \hat{\Theta}^{-1}  = -\hat{J}_{z}$ and $\hat{\Theta}$ changes $i \rightarrow -i$, therefore, we have $[\hat{\Theta}, \hat{R}_{z}(\pi)] = 0$. 

Finally, we move to the commutation relation between $\hat{\Theta} $ and $\hat{Q}(T/2)$. Although $\hat{\Theta}$ and $\hat{Q}(T/2)$ fulfill the relation $\hat{\Theta} \hat{Q}(T/2) = \hat{Q}(-T/2) \hat{\Theta}$, indicating $\hat{\Theta}$ does not commute with $\hat{Q}(T/2)$ in general, we note that since every FM is periodic in time as $| \Phi_{\alpha}(t)\rangle \rangle = | \Phi_{\alpha}(t +T)\rangle \rangle$, it then gives rise to 
\begin{equation}
 \hat{\Theta} \hat{Q}(T/2) | \Phi_{\alpha}(t)\rangle \rangle = \hat{\Theta} \hat{Q}(-T/2) | \Phi_{\alpha}(t)\rangle \rangle  = \hat{Q}(T/2) \hat{\Theta} | \Phi_{\alpha}(t)\rangle \rangle.
\end{equation}
Thus, in terms of the FMs, $\hat{\Theta} $ commutes with $\hat{Q}(T/2)$ as well.

\section{ Related properties for $P_{\alpha}(\phi)$} \label{Appendix_2}
We derive here the related properties of $P_{\alpha}(\phi)$ presented in Sec. \ref{API_few}. Before proceeding, let us first point out two preliminaries: the former unveils the relation between two FMs under the $S_{\phi}^{I}$ transformation [c.f. Eq.\eqref{S_P_I}] and the latter reveals an interesting property for the ACS. 

Followed by the discussions in Sec. \ref{API_two_phi}, for two Floquet Hamiltonians $\hat{H}_{F1}(t) = \hat{H}_{F}(t, \phi)$ and $\hat{H}_{F2}(t) = \hat{H}_{F}(t, -\phi)$ that are related by the symmetry operator $\hat{S}_{\phi}^{I} = \hat{R}_{y}(\pi) \otimes \hat{\Theta}$, their FMs $| \Phi_{\alpha}^{\phi}(t) \rangle \rangle$ and $| \Phi_{\alpha}^{-\phi}(t) \rangle \rangle$ satisfy $| \Phi_{\alpha}^{\phi}(t) \rangle \rangle =  \eta \hat{S}_{\phi}^{I} | \Phi_{\alpha}^{-\phi}(t) \rangle \rangle$, with $\eta$ being an arbitrary phase factor. Accordingly, the corresponding expansion coefficients for  $| \Phi_{\alpha}^{\phi}(t) \rangle \rangle$ and $| \Phi_{\alpha}^{-\phi}(t) \rangle \rangle$ fulfill the relation 
\begin{equation}
C^{\alpha, \phi}_{m,n} = [C^{\alpha, -\phi}_{m,n}]^{\ast}. \label{cof_phi_minus_phi}
\end{equation}
Eq.\eqref{cof_phi_minus_phi} can be roughly understood as follows: since the time-reversal operator $\hat{\Theta}$ represents a joint operation consisting of a complex conjugation and a spatial rotation of $\pi$ about the $y$-axis, the additional rotation $\hat{R}_{y}(\pi) $ results in the net effect for the $\hat{S}_{\phi}^{I} $ operator being a complex conjugation. For the use in the discussions of $P_{\alpha}(\phi)$ below,  we further set $t = 0$ for the FMs, thus it gives rise to
\begin{align}
| \Phi_{\alpha}^{\phi}(0) \rangle &= \sum_{m,n} C_{m,n}^{\alpha,\phi} | l,m \rangle = \sum_m D_m^{\alpha,\phi} | l,m \rangle, \nonumber \\
| \Phi_{\alpha}^{-\phi}(0) \rangle &= \sum_{m,n} C_{m,n}^{\alpha,-\phi} | l,m \rangle = \sum_{m,n} [C_{m,n}^{\alpha,\phi} ]^{\ast}| l,m \rangle = \sum_m [D_m^{\alpha,\phi}]^{\ast} | l,m \rangle,
\end{align}
with $D_m^{\alpha,\phi}  = \sum_{n}C_{m,n}^{\alpha,\phi} $. Here we note that at $t = 0$ the FM is solely defined in the Hilbert space $\mathcal{R} $, which allows for the expression $| \Phi_{\alpha}^{\phi}(0) \rangle$ [$| \Phi_{\alpha}^{-\phi}(0) \rangle$], instead of using the double bracket.

Next, we illustrate an interesting property for the ACS. Based on the form written in the Eq.\eqref{ACS_spin}, for the associated wavefunction, defined as $\psi_{m}(\theta, \varphi )  = \langle l,m |\theta, \varphi \rangle$, it follows that 
\begin{equation}
\psi_{m} (\theta, \varphi ) = \psi_{m}^{\ast} (\theta, -\varphi ).  \label{cof_PR_FM_even}
\end{equation}
Since $\psi_{m} (\theta, \varphi ) $ is periodic in $\varphi$ with period $2\pi$, we immediately notice that $\psi_{m} (\theta, \varphi ) = \psi_{m}^{\ast} (\theta, \varphi)$ only for the case $ \varphi = \pi$.

Equipped with the above knowledge, let us first demonstrate the relations $P_{\alpha}(\phi) = P_{\alpha}(-\phi) $ for $|\Psi(0) \rangle = |\theta,  \varphi = \pi \rangle$ and $P_{\alpha}(\phi)  \neq P_{\alpha}(-\phi) $ for $|\Psi(0) \rangle = |\theta,  \varphi \neq \pi \rangle$, which accounts for the symmetry-breaking phenomena observed in $\overline{J}_{z}(\phi)$. Since for $|\Psi(0) \rangle = |\theta, \pi \rangle$, we have
\begin{align}
P_{\alpha}(\phi) & =  \langle  \Phi_{\alpha}^{\phi}(0) |\theta, \varphi \rangle \langle \theta, \varphi|  \Phi_{\alpha}^{\phi}(0) \rangle \nonumber \\
& = \sum_{m,m^{\prime}} [D_m^{\alpha,\phi} ]^{\ast}  \psi_{m}(\theta, \varphi ) \times  D_{m^{\prime}}^{\alpha,\phi}  \psi_{m^{\prime}}^{\ast}(\theta, \varphi ),
\end{align}
and 
\begin{align}
P_{\alpha}(-\phi) & =  \langle  \Phi_{\alpha}^{-\phi}(0) |\theta, \varphi \rangle \langle \theta, \varphi|  \Phi_{\alpha}^{-\phi}(0) \rangle \nonumber \\
& = \sum_{m,m^{\prime}} D_m^{\alpha,\phi}   \psi_{m}(\theta, \varphi )  \times  [D_{m^{\prime}}^{\alpha,\phi}]^{\ast} \psi_{m^{\prime}}^{\ast}(\theta, \varphi ).
\end{align}
It immediately indicates $P_{\alpha}(\phi) = P_{\alpha}(-\phi) $ which holds only for the case $\psi_{m} (\theta, \varphi ) = \psi_{m}^{\ast} (\theta, \varphi )$, therefore, for the initial state $|\Psi(0) \rangle = |\theta, \varphi = \pi \rangle$. 

In a similar way, the symmetry $[P_{\alpha}(\phi)]_{\varphi}  = [P_{\alpha}(-\phi)]_{-\varphi}$ in Sec.\ref{API_few} can be proven as 
\begin{align}
[P_{\alpha}(-\phi)]_{-\varphi} & =  \langle  \Phi_{\alpha}^{-\phi}(0) |\theta, -\varphi \rangle \langle \theta, -\varphi|  \Phi_{\alpha}^{-\phi}(0) \rangle \nonumber \\
& = \sum_{m,m^{\prime}} D_m^{\alpha,\phi}  \psi_{m}(\theta, -\varphi )  \times  [D_{m^{\prime}}^{\alpha,\phi}]^{\ast} \psi_{m^{\prime}}^{\ast}(\theta, -\varphi ) \nonumber \\
& = \sum_{m,m^{\prime}} D_m^{\alpha,\phi}  \psi_{m}^{\ast}(\theta, \varphi )  \times  [D_{m^{\prime}}^{\alpha,\phi}]^{\ast} \psi_{m^{\prime}}(\theta, \varphi ) \nonumber \\
&= [P_{\alpha}(\phi)]_{\varphi}, \label{two_diff_P}
\end{align}
where we have employed the relation $\psi_{m} (\theta, \varphi ) = \psi_{m}^{\ast} (\theta, -\varphi )$ in Eq.\eqref{cof_PR_FM_even}. Let us note again that the above equation explains the symmetry relation for $\overline{J}_{z}(\phi)$ obtained in Eq.\eqref{symmetry_Jz_two_initial}.

\section{Determining the API via the TAHD} \label{Appendix_3}
Finally, we demonstrate how the API $\overline{J_{z} }$ can be calculated in terms of the TAHD $\overline{Q}_{H} (\theta, \varphi)$ as expressed in Eq.\eqref{API_husimi}. Followed by the works in Refs \cite{ACS_1, ACS_TAHD_1,ACS_TAHD_2,ACS_TAHD_3}, we introduce two functions $Q_{A}(\theta, \varphi) $ and $P_{A}(\theta, \varphi) $ corresponding to an arbitrary operator $\hat{A}$ with 
\begin{equation}
Q_{A}(\theta, \varphi) = \langle \theta, \varphi| \hat{A} | \theta, \varphi \rangle
\end{equation}
and $P_{A}(\theta, \varphi)$ is defined in the integral form 
\begin{equation}
\hat{A} = \frac{2l+1}{4\pi} \int d\Omega ~ P_{A}(\theta, \varphi)~ | \theta, \varphi \rangle \langle \theta, \varphi |,
\end{equation}
with $l = N/2$ being the quantum number for the total angular momentum. Due to the over-completeness property for the ACS, the expectation value for $\hat{A}$ can be expressed as \cite{ACS_TAHD_1}
\begin{equation}
\langle \hat{A} \rangle = \text{Tr} (\hat{\rho} \hat{A}) =  \int d\Omega ~Q_{H}(\theta, \varphi) P_{A}(\theta, \varphi),
\end{equation}
with $\hat{\rho}$ being the system's density matrix and $Q_{H}(\theta, \varphi)$ is the Husimi distribution given in Eq.\eqref{husimi_t} for a fixed time $t$.
In this way, the API $\overline{J_{z} }$ can be formulated as 
\begin{align}
\overline{J_{z} } &= lim_{\tau,  \tau^{\prime} \rightarrow \infty} ~\frac{2}{N  \tau^{\prime} } \int_{\tau}^{\tau + \tau^{\prime}} dt~ \langle \hat{J}_{z}\rangle(t) \nonumber \\
&= lim_{\tau,  \tau^{\prime} \rightarrow \infty} ~\frac{1}{ \tau^{\prime} } \int_{\tau}^{\tau + \tau^{\prime}} dt \int d\Omega ~Q_{H} (\theta, \varphi,t) \left[P_{J_{z}}(\theta, \varphi)/l \right]  \nonumber \\
&= \int d \Omega ~\overline{Q}_{H} (\theta, \varphi)  \overline{P}_{J_{z}}(\theta, \varphi), 
\end{align}
with $\overline{P}_{J_{z}}(\theta, \varphi) = P_{J_{z}}(\theta, \varphi)/l $ and $P_{J_{z}}(\theta, \varphi)  = (l+1) \text{cos}\theta$ \cite{ACS_TAHD_2}. For the many-particle regime ($l \gg 1$), we have $\overline{P}_{J_{z}}(\theta, \varphi) \approx \text{cos}\theta = J_{z}(\theta, \varphi)$. This demonstrates the validity of Eq.\eqref{API_husimi}.


\begin{thebibliography}{10}
\bibitem {cold_atom_rev} W. D. Phillips, Rev. Mod. Phys. {\bf70}, 721 (1998); I. Bloch, J. Dalibard, and W. Zwerger, Rev. Mod. Phys. {\bf 80}, 885 (2008).
\bibitem{BEC_1} M. H. Anderson, J. R. Ensher, M. R. Matthews, C. E. Wieman, and E. A. Cornell, Science {\bf 269}, 198 (1995).
\bibitem{BEC_2} K. B. Davis, M.-O. Mewes, M. R. Andrews, N. J. van Druten, D. S. Durfee, D. M. Kurn, and W. Ketterle, Phys. Rev. Lett. {\bf75}, 3969 (1995).
\bibitem {BH_exp_1} M. Greiner, O. Mandel, T. Esslinger, T. W. H\"ansch, and I. Bloch, Nature (London) {\bf415}, 39 (2002).
\bibitem {BH_exp_2} I. B. Spielman, W. D. Phillips, and J. V. Porto, Phys. Rev. Lett. {\bf98}, 080404 (2007).
\bibitem {BH_exp_3} T. St\"oferle, H. Moritz, C. Schori, M. K\"ohl, and T. Esslinger, Phys. Rev. Lett. {\bf92}, 130403 (2004).

\bibitem{mixture_exp_ff_1} C. A. Regal, M. Greiner, and D. S. Jin, Phys. Rev. Lett. {\bf 92}, 040403 (2004).
\bibitem{mixture_exp_ff_2} M. W. Zwierlein, C. A. Stan, C. H. Schunck, S. M. F. Raupach, A. J. Kerman, and W. Ketterle, Phys. Rev. Lett. {\bf 92}, 120403 (2004).

\bibitem{DW_exp_1} M. R. Andrews, C. G. Townsend, H.-J. Miesner, D. S. Durfee, D. M. Kurn, and W. Ketterle, Science {\bf275}, 637 (1997).
\bibitem{DW_exp_2} A. Rohrl, M. Naraschewski, A. Schenzle, and H. Wallis, Phys. Rev. Lett. {\bf78}, 4143 (1997).
\bibitem{DW_exp_3} M. Albiez, R. Gati, J. F\"olling, S. Hunsmann, M. Cristiani, and M. K. Oberthaler, Phys. Rev. Lett. {\bf95}, 010402 (2005).

\bibitem{BJJ_1} B. D. Josephson, Phys. Lett. {\bf1A}, 251 (1962).
\bibitem{BJJ_2} R. Gati and M. K. Oberthaler, J. Phys. B: At. Mol. Opt. Phys. {\bf40} R61 (2007).
\bibitem{BJJ_Rabi_1} A. Smerzi, S. Fantoni, S. Giovanazzi, and S. R. Shenoy, Phys. Rev. Lett. {\bf79}, 4950 (1997).
\bibitem{BJJ_Rabi_2} S. Raghavan, A. Smerzi, S. Fantoni, and S. R. Shenoy, Phys. Rev. A {\bf59}, 620 (1999).
\bibitem{BJJ_Rabi_3} G. J. Milburn, J. Corney, E. M. Wright, and D. F. Walls, Phys. Rev. A {\bf55}, 4318 (1997).
\bibitem{BJJ_Frag_1} K. Sakmann, A. I. Streltsov, O. E. Alon, and L. S. Cederbaum, Phys. Rev. A {\bf89}, 023602 (2014).
\bibitem{BJJ_Frag_2} K. Sakmann, A. I. Streltsov, O. E. Alon, and L. S. Cederbaum, Phys. Rev. A {\bf82}, 013620 (2010).
\bibitem{BJJ_Squeeze_1} J. Est\`eve, C. Gross, A. Weller, S. Giovanazzi, and M. K. Oberthaler, Nature (London) {\bf455}, 1216 (2008).
\bibitem{BJJ_Squeeze_2} B. Juli\'a-D\'iaz, T. Zibold, M. K. Oberthaler, M. Mel\'e-Messeguer, J. Martorell, and A. Polls, Phys. Rev. A {\bf86}, 023615  (2012).
\bibitem{BJJ_Few_1}  S. Z\"ollner, H.-D. Meyer, and P. Schmelcher, Phys. Rev. Lett. {\bf100}, 040401 (2008).
\bibitem{BJJ_Few_2} B. Chatterjee, I. Brouzos, S. Z\"ollner, and P. Schmelcher, Phys. Rev. A {\bf82}, 043619 (2010).
\bibitem{BJJ_Few_3} S. Z\"ollner, H.-D. Meyer, and P. Schmelcher, Phys. Rev. A {\bf78}, 013621 (2008).
\bibitem{BJJ_Few_4} S. Z\"ollner, H.-D. Meyer, and P. Schmelcher, Phys. Rev. A {\bf74}, 063611 (2006).
\bibitem{BJJ_Few_5} S. Z\"ollner, H.-D. Meyer, and P. Schmelcher, Phys. Rev. A {\bf74}, 053612 (2006).

\bibitem{Driven_rev_1}  M. Grifoni, P. H\"anggi, Physics Reports {\bf304},  229 (1998).
\bibitem{Driven_rev_2} S. Kohler , J. Lehmann, P. H\"anggi, Physics Reports {\bf406},  379 (2005).
\bibitem{Driven_rev_3}  A. Eckardt, Rev. Mod. Phys. {\bf89}, 011004  (2017).

\bibitem{Driven_BEC_Mott_1} K. W. Madison, M. C. Fischer, R. B. Diener, Q. Niu, and M. G. Raizen, Phys. Rev. Lett. {\bf81}, 5093 (1998).
\bibitem{Driven_BEC_Mott_2} A. Eckardt, C. Weiss, and M. Holthaus, Phys. Rev. Lett. {\bf95}, 260404 (2005).
\bibitem{Driven_BEC_Mott_3} A. Zenesini, H. Lignier, D. Ciampini, O. Morsch, and E. Arimondo, Phys. Rev. Lett. {\bf102}, 100403  (2009).

\bibitem{CDT_1} F. Grossmann, T. Dittrich, P. Jung, and P. H\"anggi, Phys. Rev. Lett. {\bf67}, 516 (1991).
\bibitem{CDT_2} J. Gong, L. Morales-Molina, and P. H\"anggi, Phys. Rev. Lett. {\bf103}, 133002 (2009).

\bibitem{Ratchet_rev} S. Denisov, S. Flach, P. H\"anggi, Physics Reports {\bf538},  77 (2014).

\bibitem{Ratchet_1} V. Lebedev and F. Renzoni, Phys. Rev. A {\bf80}, 023422 (2009).
\bibitem{Ratchet_2}  M. Schiavoni, L. Sanchez-Palencia, F. Renzoni, and G. Grynberg, Phys. Rev. Lett. {\bf90}, 094101 (2003).
\bibitem{Ratchet_3} T. Salger, S. Kling, T. Hecking, C. Geckeler, L. Morales-Molina, M. Weitz, Science {\bf 326}, 1241 (2009).
\bibitem{Ratchet_4} M. Brown and F. Renzoni, Phys. Rev. A {\bf77}, 033405 (2008).


\bibitem{Ratchet_5} P. Reimann, M. Grifoni, and P. H\"anggi, Phys. Rev. Lett. {\bf79}, 10 (1997).
\bibitem{Ratchet_6} S. Flach, O. Yevtushenko, and Y. Zolotaryuk, Phys. Rev. Lett. {\bf84}, 2358 (2000).
\bibitem{Ratchet_7} H. Schanz,  M.-F. Otto,  R. Ketzmerick, and T. Dittrich, Phys. Rev. Lett. {\bf87}, 070601 (2001).
\bibitem{Ratchet_8} T. S. Monteiro, P. A. Dando, N. A. C. Hutchings, and M. R. Isherwood, Phys. Rev. Lett. {\bf89}, 194102 (2002).
\bibitem{Ratchet_9} S. Denisov, L. Morales-Molina, S. Flach, and P. H\"anggi, Phys. Rev. A {\bf75}, 063424 (2007).
\bibitem{Ratchet_10} C. E. Creffield and F. Sols, Phys. Rev. Lett. {\bf103}, 200601 (2009).
\bibitem{Ratchet_11} T. Wulf, C. Petri, B. Liebchen, and P. Schmelcher, Phys. Rev. E {\bf90}, 042913 (2014).
\bibitem{Ratchet_12} A. K. Mukhopadhyay, B. Liebchen, T. Wulf, and P. Schmelcher, Phys. Rev. E {\bf93}, 052219 (2016).

\bibitem{Ratchet_13} S. Matthias and F. M\"uller, Nature  {\bf424}, 53 (2003).
\bibitem{Ratchet_14} A. K. Mukhopadhyay, B. Liebchen, and P. Schmelcher, Phys. Rev. Lett. {\bf120}, 218002 (2018).
\bibitem{Ratchet_15} J. F. Wambaugh, C. Reichhardt, and C. J. Olson, Phys. Rev. E {\bf65}, 031308 (2002).
\bibitem{Ratchet_16} C. Petri, F. Lenz, B. Liebchen, F. Diakonos and P. Schmelcher, Europhys. Lett. {\bf95}, 30005 (2011).
\bibitem{Ratchet_17} T. Wulf, C. Petri, B. Liebchen, and P. Schmelcher, Phys. Rev. E {\bf86}, 016201 (2012).

\bibitem{BJJ_Driven_1} M. Holthaus and S. Stenholm, Eur. Phys. J. B {\bf20}, 451 (2001).
\bibitem{BJJ_Driven_2} G. Watanabe and H. M\"akel\"a, Phys. Rev. A {\bf85}, 053624 (2012).

\bibitem{kicked_top} F. Haake, M. Kus and R. Scharf, Z. Phys. B {\bf65}, 381 (1987).

\bibitem{Feshbach_0} A. C. Pflanzer, S. Z\"ollner, and P. Schmelcher, Phys. Rev. A {\bf81}, 023612 (2010).
\bibitem{Feshbach_1} M. Olshanii, Phys. Rev. Lett. {\bf 81}, 938 (1998).
\bibitem{Feshbach_2} C. Chin, R. Grimm, P. Julienne, and E. Tiesinga, Rev. Mod. Phys. {\bf 82}, 1225 (2010).
\bibitem{Feshbach_3} T. K\"ohler, K. G\'oral, and P. S. Julienne, Rev. Mod. Phys. {\bf 78}, 1311 (2006).

\bibitem{Floquet_1} M. Holthaus, J. Phys. B: At. Mol. Opt. Phys. {\bf49}, 013001 (2016).

\bibitem{Non_degenerate}  J. von Neumann, E. Wigner, Phys. Z. {\bf30}, 467 (1929).
\bibitem {QM_Sakurai} J. J. Sakurai  and J. Napolitano, \textit{Modern Quantum Mechanics}, (Cambridge University Press, Cambridge, UK, 2017).
\bibitem {GPE_1} C. J. Pethick and H. Smith, \textit{Bose-Einstein Condensation in Dilute Gases}, (Cambridge University Press, New York, 2008).

\bibitem {PS}M. Tabor, \textit{Chaos and Integrability in nonlinear Dynamics: An Introduction}, (Wiley-Interscience, USA, 1989).


\bibitem{ACS_1} F. T. Arecchi, Eric Courtens, R. Gilmore, and H. Thomas, Phys. Rev. A {\bf6}, 2211 (1972).
\bibitem{ACS_2} J. M. Radcliffe, J. Phys. A: Gen. Phys. {\bf4}, 313 (1971).
\bibitem{ACS_3} T. Zibold, E. Nicklas, C. Gross, and M. K. Oberthaler, Phys. Rev. Lett. {\bf105}, 204101 (2010).
\bibitem{ACS_4} J. Tomkovic, W. Muessel, H. Strobel, S. L\"ock, P. Schlagheck, R. Ketzmerick, and M. K. Oberthaler, Phys. Rev. A {\bf95}, 011602(R) (2017).


\bibitem {GPE_2} M. Greiner, O. Mandel, T. Esslinger, T. W. H\"ansch and I. Bloch, Nature {\bf 415}, 39 (2002).


\bibitem {dma1_1} O. Penrose and L. Onsager, Phys. Rev. {\bf104}, 576 (1956).
\bibitem {dma1_2} K. Sakmann, A. I. Streltsov, O. E. Alon, and L. S. Cederbaum, Phys. Rev. A {\bf78}, 023615 (2008).

\bibitem {TAHD_1}  K. Husimi, Proc. Phys. Math. Soc. Jpn. {\bf22}, 264 (1940).
\bibitem {TAHD_2}  A. Piga, M. Lewenstein, and J. Q. Quach, Phys. Rev. E {\bf99}, 032213 (2019).

\bibitem {ACS_TAHD_1} B. Siram Shastry, G. S. Argarwal, and I. Rana, Pramana  {\bf11}, 85 (1978).
\bibitem {ACS_TAHD_2} E. H. Lieb, Commun. Math. Phys. {\bf31}, 327 (1973).
\bibitem {ACS_TAHD_3} R. Gilmore,  J. Phys. A: Math. Gen. {\bf9}, L65 (1976).
\end{thebibliography}
\end{document}